%% file: elsarticle-template-num.tex
\documentclass[preprint,12pt]{elsarticle}

\usepackage{amssymb}
\usepackage{url}
\usepackage{xcolor}
\usepackage{caption}
\usepackage{subcaption}
\newcommand{\Name}{SysPro}
\newcommand{\ignore}[1]{}

\begin{document}

\begin{frontmatter}

\title{SysPro: Reproducing System-level Concurrency Bugs from Bug Reports}

\author[inst1]{Tarannum Shaila Zaman}
\affiliation[inst1]{organization={Department of Information Systems, University of Maryland Baltimore County},
            addressline={Baltimore}, 
            postcode={21250}, 
            state={Maryland},
            country={USA}}
\author[inst2]{Zhihui Yan}
\affiliation[inst2]{organization={Zhejiang Provincial Financial Holdings Co,ltd},
            addressline={Hangzhou}, 
            country={China}}

\author[inst3]{Chen Wang}
\affiliation[inst3]{organization={The University of Melbourne},
            addressline={Parkville}, 
            country={Australia}}

\author[inst4]{Chadni Islam}
\affiliation[inst4]{organization={School of Computer Science, Queensland University of Technology (QUT)},
            addressline={Brisbane}, 
            country={Australia}}

\author[inst5]{Jiangfan Shi}
\affiliation[inst5]{organization={Zhejiang University},
            addressline={Hangzhou}, 
            country={China}}

\author[inst6]{Tingting Yu\corref{cor1}}
\affiliation[inst6]{organization={Department of Computer Science and Engineering, University of Connecticut},
            addressline={Storrs}, 
            postcode={06269-4155}, 
            state={Connecticut},
            country={USA}}

            \cortext[cor1]{Corresponding author. Part of the work was conducted at University of Kentucky.}

\begin{abstract}
Reproducing system-level concurrency bugs requires both input data and the precise interleaving order of system calls. This process is challenging because such bugs are non-deterministic, and bug reports often lack the detailed information needed. Additionally, the unstructured nature of reports written in natural language makes it difficult to extract necessary details. Existing tools are inadequate to reproduce these bugs due to their inability to manage the specific interleaving at the system call level. To address these challenges, we propose \Name{}, a novel approach that automatically extracts relevant system call names from bug reports and identifies their locations in the source code. It generates input data by utilizing information retrieval, regular expression matching, and the category-partition method. This extracted input and interleaving data are then used to reproduce bugs through dynamic source code instrumentation. Our empirical study on real-world benchmarks demonstrates that \Name{} is both effective and efficient at localizing and reproducing system-level concurrency bugs from bug reports.
\end{abstract}


\end{frontmatter}


\input{introduction}

\input{motivation}

\input{Approach}

\input{experiment}

\input{discussion}

\input{relatedwork}
\input{conclusion}
\section{CRediT authorship contribution statement}
\textbf{Tarannum Shaila Zaman:} Conceptualization, Data curation, Investigation, Methodology, Software, Validation, Writing – original draft, Writing – review \& editing
\textbf{Tingting Yu}: Writing – review \& editing, Validation, Supervision.
\section{Declaration of competing interest}
The authors declare that they have no known competing financial interests or personal relationships that could have appeared to influence the work reported in this paper.
\section{Data availability}
I have included the link to my supplementary material (which
includes the code and data used) in the following link: \url{https://github.com/tarannumzaman/ReproSys/tree/main}
\section{Acknowledgments}
Part of this work was conducted at the University of Kentucky. This work was supported in part by NSF grants CCF-2348277 and CCF-2518445.

 \bibliographystyle{elsarticle-num} 
 \bibliography{sample-base}

\end{document}

%% file: Introduction.tex
\section{Introduction}
The increasing use of multi-core platforms, interrupt-driven devices, and distributed systems has made concurrent systems prevalent. Despite their performance benefits, guaranteeing the correctness of concurrent programs is difficult due to non-deterministic behavior and complex execution interleaving \cite{DBLP:conf/sigsoft/BianchiPT17}. 
Reproducing and localizing concurrency faults is a time-consuming endeavor,  \cite{1342780, 10971715} and challenging. Unlike traditional software bugs, reproducing or localizing system-level concurrency bugs requires both input information and the interleaving details of system calls.

Modern software projects use bug-tracking systems like GNU \cite{GNU}, Bugzilla \cite{redhat}, and Debian \cite{debian} to expedite the resolution of concurrency bugs. These systems allow testers and users to report issues they have identified in an application \cite{recdroid, 10.1145/3106237.3106285, 10.1145/3238147.3238204, 6704866}. However, unstructured and ambiguous information often makes bug reports difficult to use. Our study of over 1,200 system-level bug reports identifies several challenges in reproducing these bugs. The major challenges include: i) Many bug reports lack essential system call names or input details required for reproduction, ii) Lexical gaps exist between human-written descriptions and technical system call names, and iii) Irrelevant system call names in some bug reports complicate targeted instrumentation. Even when developers fully understand a bug report, reproducing the issue remains challenging because concurrent programs behave differently in the same environment across different executions.

Existing work (\cite{7774517, DBLP:conf/sigsoft/YuZW17}) for reproducing system-level concurrent requires console logs or human intervention to reproduce this kind of bug. For instance, Descry \cite{DBLP:conf/sigsoft/YuZW17} utilizes default console logs for bug reproduction, which may not be available in real-world scenarios where only a bug report with a bug description exists. Another approach, RRF \cite{7774517}, requires human intervention to collect input and interleaving information from the bug reports.  
Thus, there is a need to develop an automated bug reproduction tool that can reproduce a bug only by analyzing the bug reports. 

In this work, we introduce \emph{\Name{}}, an effective bug reproduction tool designed to assist developers in reproducing and pinpointing the causes of concurrency failures within specific application functions using only bug reports. \emph{\Name{}} targets inter-process (system-level) concurrency bugs. These system-level concurrency bugs arise when two processes access a shared resource, potentially in a different order than originally intended. These bugs can corrupt the persistent storage and other system-wide resources \cite{Simracer} that can result in system-wide crashes. In contrast, intra-process (thread-level) concurrency faults typically affect volatile memory within a single process \cite{DBLP:conf/sigsoft/YuZW17}. Research indicates that over 73\% of the reported race conditions in Linux distributions are system-level races \cite{10.1145/2043556.2043589}. To reproduce a system-level concurrency bug, developers require not only input data but also specific interleaving orders at the system call level. Our work focuses on the extraction of such interleaving information from bug reports, specifically, a sequence of system calls on the same shared resource from the same or different processes.

Our goal is to analyze the application bug report to identify relevant system calls and functions leading to the failure in the source code and use the identified information to guide the instrumentation of the program. The instrumented system calls are permuted at runtime until the reported bug is successfully reproduced. Our hypotheses include: 1) bug reports contain a specific set of vocabulary (system call names, function names, etc.) related to buggy interleavings that make the automated text extraction and mapping possible; 2) the extracted information focuses on a small set of functions and system calls, which can narrow down the scope of instrumentation and thus speeds up the bug reproduction.

\Name{} takes as input a bug report that describes a system-level concurrency bug. If the bug is successfully reproduced, it outputs an interleaving schedule consisting of system calls and their associated function names and locations in the program that cause the failure. \Name{} works in two major phases. In the first phase, \Name{} extracts system call names potentially causing the bug from the bug report and maps them to the application source code. In this phase, \Name{} uses Natural Language Processing (NLP) and data mining techniques to analyze bug reports. It then performs static analysis on the source code to use a structured information retrieval technique to calculate the similarity between the bug report and the source code. In the second phase, \Name{} reproduces the bug based on the extracted information. Specifically, \Name{} first generates inputs from the bug reports with the help of regular expression matching and category partition method \cite{10.1145/62959.62964}. It then uses the input information and the interleaving information to reproduce the bug by dynamic binary instrumentation \cite{Pin_A_Dynamic_Binary_Instrumentation_Tool} of the source code.

\Name{} is implemented as a software tool using the \emph{srcML} static code analysis tool \cite{srcml}, \emph{NLTK} for natural language processing with Python \cite{nltk}, the \emph{Gensim} Python library \cite{gensim} for information retrieval, the \emph{SPMF} java open source data mining library \cite{SPMF}, and a dynamic binary instrumentation tool, \emph{PIN} \cite{Pin_A_Dynamic_Binary_Instrumentation_Tool}. To evaluate \Name{}, we conducted an empirical study on \emph{19} bug reports from \emph{17} applications with known reproducible real-world concurrency failures. Our results demonstrate that \Name{} effectively identifies system call names that cause buggy interleavings in the source code by using the information from the bug reports. We also found that by using the buggy interleaving information along with the extracted input, \Name{} is effective in reproducing system-level concurrency bugs automatically from bug reports.

In summary, in this paper, we make the following contributions:
\begin{itemize}
\item We conduct an empirical study on \emph{1,210} system-level concurrency bug reports to summarize their characteristics and identify three major challenges in reproducing these bugs from reports. This study guides us in developing an automated framework that reproduces real-world system-level concurrency bugs.
We propose \Name{}, a semi-automated system for reproducing system-level concurrency bugs from bug reports, requiring manual effort for test case creation when the bug report does not explicitly include the test cases.
\item We implement \Name{} and conduct an empirical study on 24 real-world bug reports, demonstrating 95.8\% effectiveness and efficiency in reproducing system-level concurrency bugs.
\end{itemize}

%% file: motivation.tex
\section{Background and Motivation}
In this section, we first discuss the basics of information retrieval-based fault localization techniques. Next, we define system-level concurrency bugs and present an example. We also include a sample bug report of this type of bug and present it as a motivating example.

Finally, we conduct an empirical study of system-level concurrency bug reports to guide the design of our approach. This study is essential because each bug report is distinct and does not follow a consistent format. Additionally, bug reports on different websites use different presentation structures. These issues lead us to perform an empirical study to gain insights into the nature of real-world system-level bug reports. Through this study, we identify the challenges of reproducing system-level bugs from bug reports and develop our \Name{} framework to address these challenges. Based on the findings, we define the problem of reproducing system-level concurrency bugs: given a bug report, a set of processes that trigger the bug, and the source code of the buggy application process mentioned in the report, we aim to replicate the failure described in the report.

\subsection{Information Retrieval-based Fault Localization}

In this section, we discuss two types of bug report-oriented fault localization techniques
based on information retrieval (IR): the basic IR approach~\cite{TFIDF, 6227210} and the structural IR approach~\cite{structuredIR, 8477206, 10.1145/508791.508919}.
In the
basic IR approach, the bug report is considered as a query and the source code files of the application are used as the search documents. The basic IR system begins with four-step prepossessing: text normalization, stop word removal, reserved word removals (for C source code), and stemming. Then we calculate the TF.IDF-based similarity score \cite{TFIDF} to find the most relevant source code files with the bug report. In the rest of the paper, where we mention the \textit{BasicIR} technique we refer this technique. 

The \textit{BasicIR} technique does not consider source code structure, i.e., each term in a source code file is considered having the same relevance with respect to the given query \cite{structuredIR}. 
Therefore, important information such as 
function names and system call names often get lost in the
relatively large number of variable names and comments 
due to the term weighting function. To overcome these challenges, researchers developed the structured IR techniques \cite{structuredIR, 8477206, 10.1145/508791.508919}. 
In this paper, we compare our work with the structured IR based technique BLUiR \cite{structuredIR}. BULiR uses two different parts of the bug report - subject and body as two queries, and four different code constructs, i) class names, ii) method names, iii) variable names, and iv) comments as document fields. In total, they perform a separate search for each of the eight combinations (query represent, document field) and then
sum document scores across all eight searches. To find a similarity score they use the TF.IDF similarity model. In Section \ref{roleofalgo}, we compare the performance of BasicIR, BLUiR, and our technique \Name{} for localizing system-level concurrency bugs.
\subsection{System-level Concurrency Bug}
\label{sysbug}
A system-level concurrency bug arises when multiple processes, signals, or interrupts access a system-wide resource (e.g., a file or device) without proper synchronization \cite{10.1145/2043556.2043589}. These resources are commonly accessed through system calls. Reproducing a system-level concurrency fault involves modeling read/write effects and synchronization operations that encompass system calls \cite{Scminer}.

We discuss a real-world system-level concurrency bug in Linux Coreutils version 6.9, reported in Bugzilla \cite{mvbug}. The bug description reveals a race condition when attempting to atomically replace the content of two files, `foo' and `bar.' The problem arises from the sequence of system calls, illustrated in Table \ref{scsequence}. By initially unlinking the target file `foo,' the \emph{mv} command introduces a race condition where `foo' may be missing when accessed by process $P_2$. This race condition violates the atomicity of the `rename()' operation. This bug persisted in Coreutils for a significant duration, affecting many programs relying on this version \cite{mvbug}.
\begin{table}[ht]
\begin{center}
\caption{Passing and Failing execution trace of \emph{mv} and \emph{cat} process} 
\label{scsequence} 
\scalebox{0.9}{
\begin{tabular}{ |p{3.75cm}|p{3.75cm}| }
\hline
\textbf{Correct Order} & \textbf{Buggy Order}\\
\hline
mv$\_$unlink(foo) & mv$\_$unlink(foo)\\
\hline
mv$\_$rename(bar,foo) & $P_2$$\_$open(foo)\\
\hline
$P_2$$\_$open(foo) & mv$\_$rename(bar,foo)\\
\hline
\end{tabular}}
\end{center}
\vspace{-2ex}
\end{table}

This bug report characterizes the file names as input data, with \emph{mv} as the command. It also contains specific system call names like \emph{unlink} and \emph{rename}, which are part of the problematic sequence leading to the bug.
To reproduce this type of bug we need two pieces of information, i) the system call names in the buggy interleaving and, ii) the input.

\textbf{Buggy Interleaving.} In this paper, we will use the term ``buggy interleaving", which is the list of system calls that are causing the bug. For this example, the system calls, \emph{unlink, open, rename} from process \emph{mv} and process \emph{$P_2$} is in the buggy interleaving. The system calls of the buggy interleavings are from two different processes. To pinpoint the bug's cause, we should focus on system calls from the \emph{mv} process, particularly the \emph{unlink} and \emph{rename} pair. These calls are the root cause of the problem. Any other system calls (e.g., `state/lstate/read') instead of \emph{open} from different processes that occur between \emph{unlink} and \emph{rename} can trigger the bug. Therefore, identifying the targeted program's system calls within the buggy interleaving is crucial to understanding and fixing the bug \cite{DBLP:conf/sigsoft/YuZW17}.

\textbf{Input.} We also need a specific input to reproduce the bug. For this example, the input is written in the bug report as ``mv bar foo" and we need to use it to reproduce the bug. This input has two parts the command part which is \textit{mv} and the data input part which is two file names, \textit{bar}, and \textit{foo}.

This bug may not be reproduced through multiple executions with the same input set. This is because we need to enforce a specific order of system calls, as outlined in Table \ref{scsequence}. This is the main challenge in reproducing these types of bugs. The process becomes even more challenging when we rely solely on the bug report to reproduce the issue. Bug reports do not always ensure that the necessary information is presented clearly. Often, they contain interleaving information, including system call names from one or more processes, which can provide valuable guidance for reproducing the bug. In the following section, we present the bug report that contains the bug described in Section \ref{sysbug} and highlight the necessary information required to reproduce the bug.

\subsection{A Motivating Example}
In this section, we present a bug report as an example.
While different bug reports pose unique challenges, as described in Section \ref{bugrepotchallenges}, we focus on a single case study here.
\ignore{
\begin{figure*}[t]
\centering
\includegraphics[width=10cm]{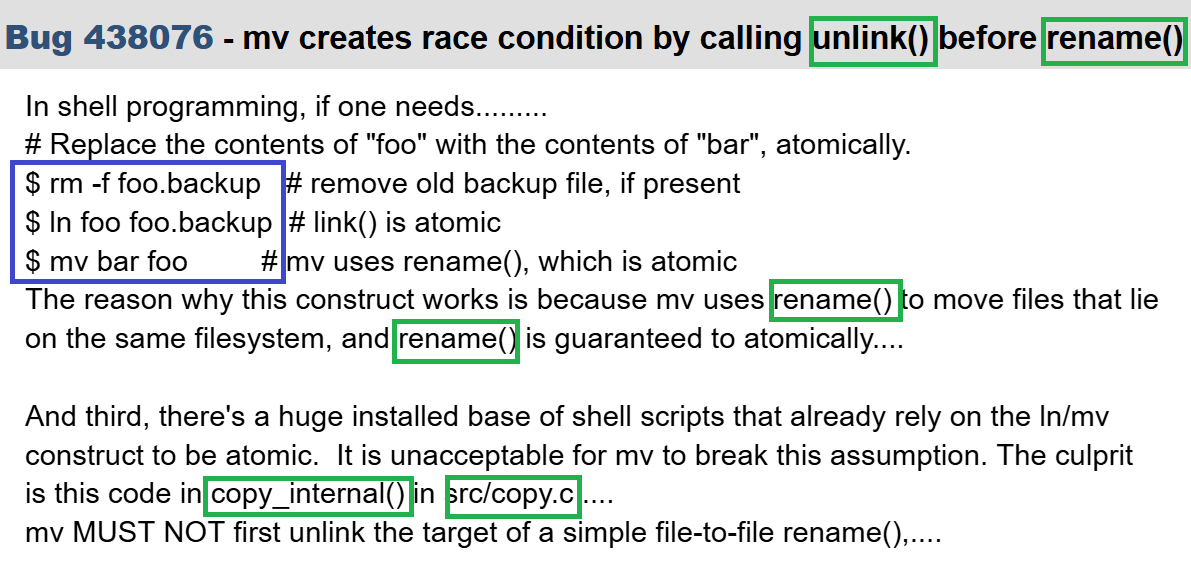}
\caption{A Part of A Sample Bug Report. \cite{mvbug}}
\label{fig:casestudybugreport}
\end{figure*}}

\begin{figure}
    \centering
    \begin{subfigure}[b]{0.4\textwidth}
        \centering
        \includegraphics[width= 7.5cm]{bugreport_example.png}
        \caption{A Part of A Sample Bug Report. \cite{mvbug}}
        \label{fig:casestudybugreport}
    \end{subfigure}
    \hfill
    \begin{subfigure}[b]{0.4\textwidth}
        \centering
        \includegraphics[width= 6.5 cm]{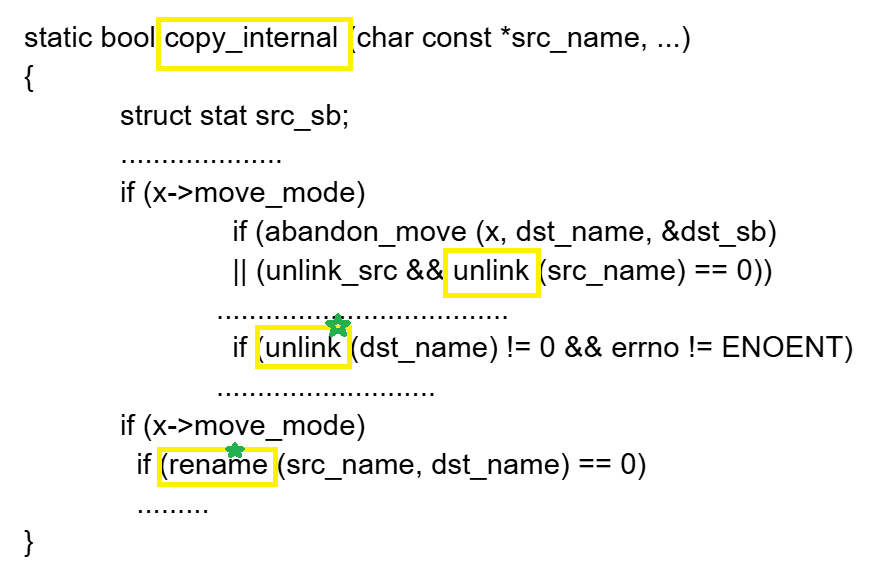}
        \caption{Code Snippet from copy.c File \cite{coreutils}.}
        \label{fig:casestudycode}
    \end{subfigure}
    \caption{An example of system-level bug report and associated source code}   
    \label{fig:example}
\end{figure}

Figure \ref{fig:casestudybugreport} presents a portion of a sample bug report \cite{mvbug} related to the mv race condition bug discussed in Section \ref{sysbug}. To reproduce the bug described in this report, we first need to extract the input. The section marked in blue indicates the input data within the bug report. Next, we must identify the system calls involved in the buggy interleaving and locate them in the source code. This allows us to determine the appropriate instrumentation points for reproducing the bug. The sections marked in green highlight the system call names, a function name, and the name of the file that contains these system calls and function.

This bug report contains 12 unique system calls, which are listed in Figure \ref{fig:sample_bugrepo2}. From this report, we need to identify the system calls involved in the buggy interleaving: unlink and rename. A careful reading of the bug report reveals that these two system calls appear not only in the description but also in the subject line of the report. Additionally, they occur more frequently than other system calls in the report. Furthermore, this specific bug report includes the name of the buggy file, \emph{copy.c}, and the function name, \emph{copy\_internal} (highlighted in the green box in Figure \ref{fig:casestudybugreport}). 

These findings motivates us to use the Structured IR technique from Section \ref{rankfile} to rank the source code files associated with this bug report. The file \emph{copy.c} ranks highest because the bug report explicitly mentions its name, a function from this file, and includes system calls present in both the subject and description. As a result, our Structured IR method identifies \emph{copy.c} as the most suspicious file.

Next, we determine the location of the faulty system calls within the \emph{copy.c} file. To achieve this, we identify system call pairs based on their co-occurrence within the same or different sentences in the bug report and rank them by frequency. In this example, we find that the system calls \texttt{unlink} and \texttt{rename} appear together most frequently in the bug report (three times). A detailed explanation of how frequent pairs are identified is provided in Section \ref{Approach3a1}.

Figure \ref{fig:casestudycode} shows a code snippet from the \emph{copy.c} file, highlighting three system calls—\texttt{unlink}, \texttt{unlink}, and \texttt{rename}—marked in yellow. We search for the system call pair \texttt{unlink} and \texttt{rename} within the highest-ranked source file, \emph{copy.c}, to identify their occurrences. This approach demonstrates the advantage of searching for system call pairs rather than individual system calls. A search for \texttt{unlink} alone returns 22 matches, while \texttt{rename} appears 24 times, making it difficult to pinpoint the exact locations for instrumentation. Our method resolves this challenge by identifying precise locations where the system call pair occurs sequentially. In \emph{copy.c}, we locate the exact line numbers where \texttt{unlink} and \texttt{rename} appear one after the other. As shown in Figure \ref{fig:casestudycode}, our approach, \Name{}, marks these occurrences with green stars, identifying them as the buggy system calls. We use this location as the instrumentation point for PIN.
\ignore{
\begin{figure*}[t]
\centering
\includegraphics[width=8 cm]{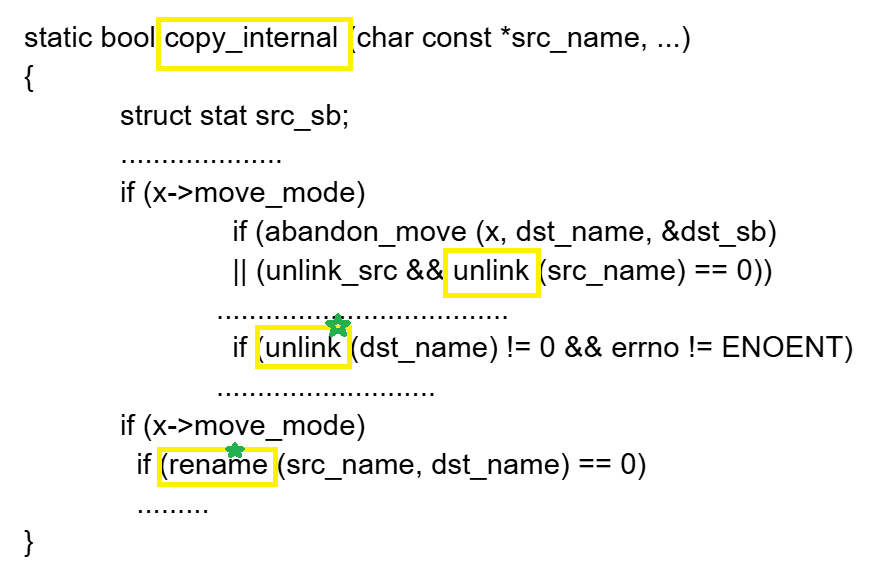}
\caption{Code Snippet from copy.c File \cite{coreutils}.}
\label{fig:casestudycode}
\end{figure*}

To reproduce the bug, we generate a test case based on the bug report. For this specific bug, we extract the test cases highlighted in blue in Figure \ref{fig:casestudybugreport} using regular expression matching. Section \ref{bugreproduce} details the process of generating inputs from bug reports.}

Finally, we use the instrumentation location and test cases generated by \Name{} to execute the buggy application. Developers run the application with PINtool for instrumentation to trigger the race condition. Previous research works \cite{Scminer, 10.1145/3476883.3520207} provide a detailed explanation of how PINtool facilitates instrumentation. Our experimental evaluation confirms that \Name{} successfully reproduces this bug in a single run by applying the extracted inputs and identified instrumentation location.

Since different bug reports have different structures and contain different types of information, we conduct an empirical study on a set of bug reports. In the following section, we examine the characteristics of bug reports related to system level concurrency issues and identify the challenges we need to address to develop an automated bug reproduction system.

\input{Study}

%% file: Study.tex
\subsection{Studying System-level 
Concurrency Bug Reports}
\label{study}
We searched various open-source bug repositories, including GNU \cite{GNU}, Bugzilla \cite{redhat}, and Debian \cite{debian}, to identify reports related to system-level concurrency bugs. We first conducted a keyword search by using keywords, such as
``race", ``concurrency", and ``process". This yielded a 
total of 5,393 bug reports.  We then manually examined these bug reports and identified
1,210 bug reports related  to system-level concurrency bugs. 

If a bug report directly mentions
bug-related system call names, indirect hints of the system call names by describing
any function/file/variable names,
or describes the action 
of a system call instead of its name,
we consider that the bug report contains the \emph{interleaving} information. 
Here, the direct information indicates that the bug report explicitly
mentions the system calls names. 
The indirect information suggests that the buggy 
system calls are implicitly mentioned in the bug report. 
Based on our study, we have identified four types of indirect information 
that is potentially useful for localizing 
 concurrency bugs in the code, including
1) natural language description of the system calls, 2) function names, 3) file names, and 4) variables. 

\begin{table}[ht]
\begin{center}
\caption{Summary of the Study on Bug Reports}
\label{studytable} 
\scalebox{.85}{
\begin{tabular}{|l|l|r|r|r|r|r|r|}
\hline
\emph{Project} & \emph{\#BR} &\multicolumn{5}{c|}{\emph{Structured Bug Report Entities}} &\emph{PName} \\
\cline{3-7}
& &\emph{Sysc.} &\emph{Desc.}&\emph{Func.} & \emph{File} & \emph{Var.} &\\ 
\hline
Bugzilla & 1011 & 35$\%$& 20$\%$& 13$\%$ & 13$\%$ & 2$\%$ & 98.5$\%$\\
\hline
Debian & 103 & 58$\%$ &12$\%$& 5$\%$ & 3$\%$ & 0$\%$ & 98$\%$\\
\hline
GNU & 96 & 52$\%$ & 20$\%$ & 27$\%$ & 10$\%$ & 10$\%$ & 97$\%$\\
\hline
Total & 1210 & 38.5$\%$ & 19 $\%$ &13$\%$ & 11.5$\%$ & 2.5$\%$ & 98$\%$ \\
\hline
\end{tabular}
}
\begin{flushleft}
\small{
$\#BR$ = number of bug reports. $Sysc.$ = System call/Signal name stated directly. $Desc.$ = Description of the System calls. $Func.$ = Function Name Stated. $File$ = File Name Stated. $Var.$ = Variable Name Stated. $PName$ = Buggy Process Name Stated.}
\end{flushleft}
\end{center}
\vspace{-2ex}
\end{table}

Table \ref{studytable} shows a summary of our study of 
1,210 bug reports. 
Column 2 shows the number of sampled bug reports selected from the searched bug reports. The next five columns show the percentage of bug reports containing system call/signal names directly, system call's description, function names, file names, 
and variable names related to the reported bug. Finally, the last column shows the percentage of bug reports that contain the buggy application name.

The overall results show that 38.5$\%$  bug reports contain the 
names of system calls or signals directly and
46$\%$ of them contain system call names in an indirect manner.
Among the indirect items, 
19$\%$ bug reports contain information about system calls in a descriptive manner, 
13$\%$ have at least one function name of the source code, 11.5$\%$
contain file names of the source code, 2.5$\%$ contain variable names of the source code, and 15.5$\%$ do not have any information about the interleavings. 98$\%$ bug reports contain the application name where the bug resides.

This study motivates us to automate the bug reproduction system and identify its associated challenges. We derive key findings that shape our framework design and rationale: (i) Most bug reports (98\%) include the name of the buggy application, so our system assumes prior knowledge of it. (ii) Many bug reports (46\%) indirectly mention system call names, so we treat these references as keywords in our structured IR system \cite{structuredIR}. In addition to the inherent challenges of this type of bug, we identify three unique challenges in reproducing these bugs solely from bug reports.

\subsection{Challenges in Reproducing System-level Bug Reports}
\label{bugrepotchallenges}
Our study of system-level concurrency bug reports highlights several challenges in identifying the relevant system calls needed to reproduce these bugs.
\noindent
\subsubsection{Lexical gaps.} 
\begin{figure}[t]
\centering
\includegraphics[scale=0.4]{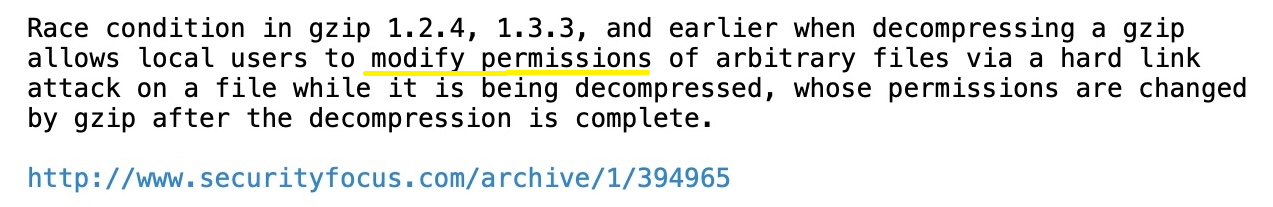}
\caption{Race condition in \texttt{gzip} program: Bugzilla \#155746}
\label{fig:sample_bugrepo1}
\vspace{-1ex}
\end{figure}

In 38\% of bug reports, system call names are not directly mentioned. For instance, as illustrated in Figure~\ref{fig:sample_bugrepo1}, the bug report~\cite{gzipbugreport} discusses the modification of file permissions, which refers to the \emph{chmod} system call. However, the actual system call name is not explicitly stated in the bug report but in the description, it states the operation of this system call.\\
\textbf{Our Solution: }To the best of our knowledge, no prior approach has directly addressed the lexical mismatch between system call names and the natural language used in bug reports. This challenge is particularly relevant for system-level concurrency bugs, where accurately identifying system call names is critical. To bridge this gap, we utilize system call manuals. By comparing the textual descriptions provided in these manuals with the content of bug reports, we identify relevant system calls. Further details of this process are provided in Section~\ref{extractsyscall}. 

\begin{figure}[t]
\centering
\includegraphics[scale=0.55]{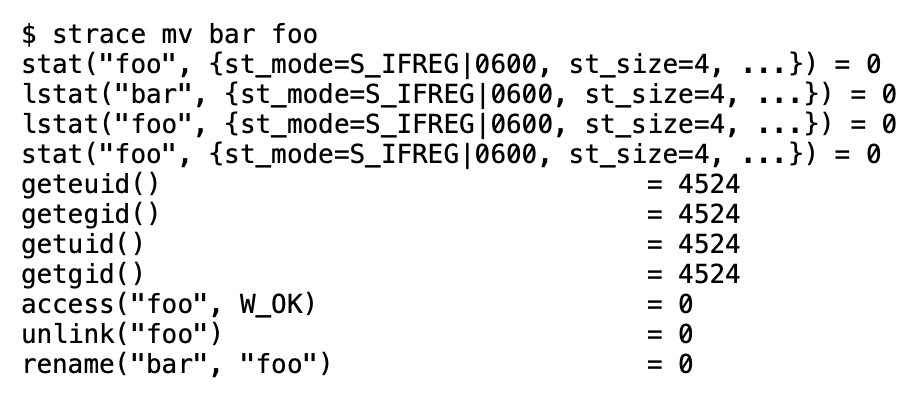}
\caption{Race condition in \texttt{mv} program: Bugzilla \#438076}
\label{fig:sample_bugrepo2}
\vspace{-2ex}
\end{figure}
\begin{figure}[b]
\centering
\includegraphics[scale=0.45]{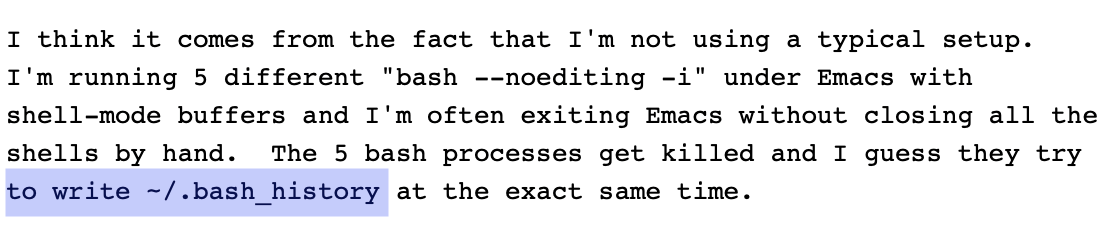}
\caption{Race condition in \texttt{bash} program: Debian \#283702}
\label{fig:sample_bugrepo3}
\end{figure}

\subsubsection{Irrelevant system calls.}We find that bug reports sometimes include multiple system call names that are unrelated to the buggy interleaving. For example, in Figure \ref{fig:sample_bugrepo2}, a bug report mentions 12 system call names in the context of the \emph{mv} program. The bug occurs when a system call, such as \emph{open} from a different process, interferes between the \emph{unlink} and \emph{rename} system calls of the \emph{mv} program. In this case, the problematic interleaving is represented as \emph{$<$ P1: unlink, P2: open, P1: rename $>$}, where ‘P1’ and ‘P2’ denote two distinct processes. However, in Figure \ref{fig:sample_bugrepo2}, only two of the twelve system calls (\emph{unlink} and \emph{rename}) are part of the buggy interleaving. This means that only 17\% of the system calls in this bug report are relevant to the buggy interleaving.\\
\textbf{Our Solution:} To extract relevant system calls, we leverage a structured Information Retrieval (IR) technique, as described in \cite{structuredIR}. Although prior research \cite{10.1109/ASE.2011.6100061, structuredIR} has shown that structured IR can improve accuracy by filtering out duplicate or irrelevant information, these works have not addressed the specific challenge of eliminating irrelevant system calls from bug reports. This challenge is particularly unique to system-level bug reports. We choose structured IR over traditional IR because it allows us to differentiate between distinct sections of a bug report—such as the subject and the body—and to assign greater importance to critical keywords \cite{10.1109/ASE.2011.6100061}.

In our approach, we create two separate query representations based on different sections of a bug report: the subject and the description. For example, in the previously mentioned \emph{mv} bug report, there are 11 system call names, two of which appear in the subject line. When applying our model, we generate separate ranking results for the subject and body queries. We then compute the average rank of the corresponding source code files. This approach enables us to emphasize essential information—particularly from the bug report subject—as further discussed in Section~\ref{rankfile}.
\subsubsection{Instrumenting target system calls.} 
To effectively reproduce bugs, it's essential to map the problematic system calls mentioned in bug reports to the corresponding source code. However, a buggy system call can often be quite common, appearing hundreds to thousands of times in the source code. For example, in Figure~\ref{fig:sample_bugrepo3}, there's a bug report related to the \emph{bash} program, where the problematic system call is 'write.' However, in the source code, this particular system call occurs more than 100 times.\\
\textbf{Our Solution: } This issue represents a unique challenge specific to system level concurrency bugs, and to the best of our knowledge, no prior work has attempted to address it automatically. To tackle this problem, we not only consider system call names but also incorporate the contextual information in which these system calls appear. The structured IR technique helps address this challenge effectively.

In our approach, we generate separate ranking results for various elements extracted from the source code, including system call names, file names, function names, and variable names. Traditional IR techniques, such as the TFIDF model \cite{TFIDF}, do not account for the structure of source code. These methods treat all terms in a source code file as equally relevant to a given query \cite{structuredIR}, which often causes important information such as function names and system call names to be overshadowed by the abundance of variable names and comments due to uniform term weighting. Our proposed model addresses this limitation by distinguishing between different code constructs. While previous works using structured IR techniques \cite{structuredIR, 6100061} have considered various contextual elements within source code, none have treated system call names as a distinct and significant context. However, for reproducing system level concurrency bugs, system call names often serve as a critical piece of information.

As per the study, 27$\%$ of the bug reports contain function names,
source file names, and variable names of the source code. 
If we can localize the function/file or variable names in the source code  
related to the buggy interleaving, we can narrow down the system call names 
by observing only those specific function/file names or the functions,
which contain the variables specified in the bug report. 
Those system calls will be ranked at the top based on
the similarity of their function/file names in the bug report. We provide a detailed explanation of this approach in Section \ref{rankfile}.\\

%% file: Approach.tex

\section{Approach}
\begin{figure*}[b]
\centering
\includegraphics[width=13.5cm]{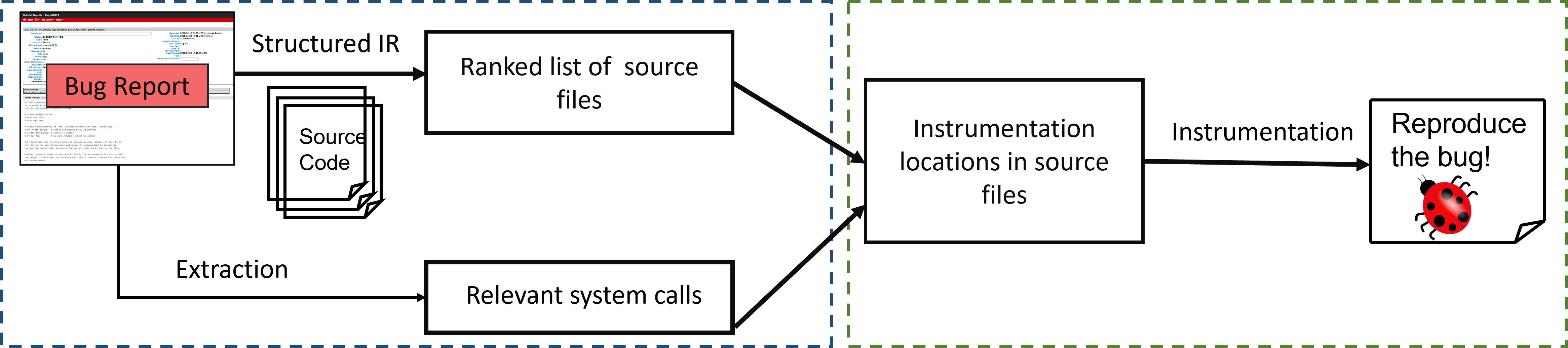}
\caption{The overview of \Name{} framework}
\label{fig:overview}
\end{figure*}

Figure \ref{fig:overview} illustrates the \Name{} approach, which consists of two major steps: 1) Localizing the system calls involving the buggy interleaving
in the target program, and 2) Reproducing the bug by instrumenting these system calls. 
In the first step, \Name{} identifies the location of 
the system calls that might be causing the bug in the source code of the target program as described in the bug report. To achieve this, \Name{} first extracts the system call names from the bug report and employs a structured information retrieval (IR)~\cite{structuredIR} technique to create a ranked list of source files that are most likely relevant to the bug report. It then pinpoints the locations of the system calls involved in the buggy interleaving in the source files. 

In the second step, \Name{} creates test cases by extracting information from the bug report. It then uses the ranked list of system call locations to instrument the application~\cite{Pin_A_Dynamic_Binary_Instrumentation_Tool}. Based on the instrumented system calls, it permutes process interleavings by controlling scheduling~\cite{Simracer}.

Among these two steps, \Name{} fully automates the first step. In the second step, it requires the user to generate test cases using the category-partition method~\cite{Ostrand88}, which relies on a Test Specification Language (TSL). However, if the bug report explicitly includes the test cases, \Name{} automatically extracts the input. Section~\ref{bugreproduce} describes this approach in detail.

\begin{figure}[t!]

{\hrule height 1.5pt}
\vspace*{3pt}
\noindent
\footnotesize{
\textbf{Algorithm for Localizing Buggy System Calls}\\
{\hrule height 0.5pt}
\vspace*{1pt}
\raggedright
1:~~\textbf{Inputs:} $BugR , SC, SysCallM$ \\
2:~~\textbf{Outputs:} $Syscall, LineNo, Function, FileName$\\
3:~~\textbf{begin}\\
4:~~\hspace{0.25cm}$KeySystemCalls$ $\leftarrow$ \texttt{ExtractSysCall} ($BugR,SysCallM$)\\
5:~~\hspace{0.25cm}$List_{{File}}$ $\leftarrow$ \texttt{TFIDFSimSrch} ($BugR$,$SC$)\\
6:~~\hspace{0.25cm}$Syscall_{pairs}$ $\leftarrow$ \texttt{CreateSyscallPairs} ($KeySystemCalls$)\\
7:~~\hspace{0.25cm}$RList_{SysCall}$ $\leftarrow$ \texttt{SearchSyscall} ($Syscall_{pairs}, RList_{File}$)\\
8:~~\textbf{Return} $Syscall, LineNo, Function, FileName$\\
}
\vspace*{1pt}
{\hrule height 1.5pt}
\caption{Localizing the buggy interleaving in the source code}
\label{algo_1st}
\end{figure}

\subsection{Localizing Buggy System Calls}
\label{approach3a}

In order to reproduce a system-level concurrency bug, it is necessary to identify the system calls involved in the buggy interleaving within the target program.

Figure \ref{algo_1st} illustrates the algorithm used to localize buggy interleavings in the source code. \Name{} takes a bug report, the source code, and the system call manual as inputs, and outputs a ranked list of potential buggy system call locations (i.e., system call name, line number, function name, and file name).

\Name{} begins by extracting system call names from the bug report and creating a list of these calls (line 4). This step is detailed in subsection \ref{extractsyscall}. Next, \Name{} utilizes a structured Information Retrieval (IR) technique to rank the source files relevant to the bug report (line 5). Unlike existing structured IR techniques~\cite{structuredIR}, \Name{} formulates different queries and structured documents by considering the unique characteristics of system-level concurrency bugs, such as system calls and their contexts (e.g., functions and variables). The top-K files are then identified as highly relevant (detailed in subsection \ref{rankfile}).

Given that the same system call may appear multiple times in the top-K files, \Name{} ranks the system calls based on their relevance to the buggy interleaving. Our approach uses the \emph{apriori} algorithm\cite{AprioriAlgo} to create system call pairs from the bug report and rank them according to their frequencies within the report (line 6). Finally, we pinpoint the locations of these system calls in the source code and identify the specific system call to be used as the instrumentation point to reproduce the bug. The process of locating the instrumentation point is described in subsection \ref{Approach3a1}.

\begin{figure}[htb]
\centering
\includegraphics[scale=0.38]{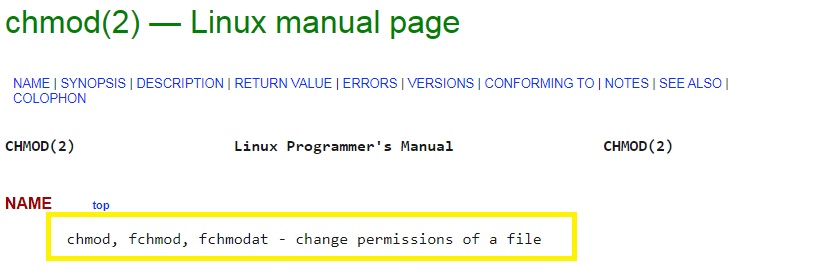}
\caption{Sample of a Linux man page for system call \textit{chmod}}
\label{fig:samplemanpage}
\end{figure}

\subsubsection{Extract System Call Names from the Bug Report}
\label{extractsyscall}
This approach begins with a simple keyword search to find system call names in the bug reports. It matches the bug report against Linux system call names \cite{Ubuntusyscall, Syscall} to identify the system calls mentioned in the bug report, referred to as \emph{KeySystemCalls}.

Our empirical study shows that, in 61$\%$ of bug reports, the system call names are not stated directly. In such cases, the \emph{KeySystemCalls} list remains empty. To address this issue, we use system call manuals to bridge the lexical gap between the system call names and the bug report. The approach creates files containing each system call name along with the description under the ``NAME" tag of the Linux manual page (Figure \ref{fig:samplemanpage}).
The approach then calculates the similarity score between the bug report and each of these files. In this process, the bug report serves as the query and all the files are the search documents. We use the TFIDF model similarity matrix \cite{TFIDF} to calculate the similarity scores. The output is a ranked list of system call names based on their scores, with higher-ranked system calls being more relevant to the reported bug. The top-n ranked system calls are saved into the \emph{KeySystemCalls} list. By default, n is set to 10, as our experimental results indicate that the most relevant system calls are typically found within the top 10 positions.

The leftmost column of Table \ref{keysyscall} shows some of the \emph{KeySystemCalls} from the bug report of \emph{mv} program~\cite{mvbug}.

\subsubsection{Locating Suspicious Source Files}
\label{rankfile}
We use structured information retrieval (IR) to generate a ranked list of source files, where the top-ranked files are more similar to the bug report and are considered to have buggy interleaving. Unlike existing work~\cite{structuredIR}, \Name{}'s structured IR considers the characteristics of system-level concurrency bugs. In our study (Section~\ref{study}), we found that important program constructs, such as system call names, their surrounding methods, and variables, are effective indicators of the locations of buggy interleaving. In our technique, we use three query fields (subject, body, and system call names) obtained from the bug report. The \emph{KeySystemCalls}, which contains the system call names of the bug report, is used as the third query. We perform three different searches for these queries where the documents are the buggy application's source files.

We use \textit{srcML} \cite{srcml} to convert the source code into an XML file and then perform static analysis on this file. With the XPath query, we list the file names, function names, variable names, and comments of the source code. Next, we create four distinct document fields. We then perform a separate search for each of the twelve (query, document field) combinations.

We use the Gensim TFIDF model and the similarity matrix \cite{TFIDF} 
to find the similarity between the bug report and the source files. 
Our implemented IR system begins with four-step prepossessing: i) text normalization, ii) stop word removal, iii) reserved word removals (for C source code), and iv) stemming. We use standard NLP \cite{nltk} prepossessing for word tokenization and stop word removal.
We perform total twelve different searches for three different queries (subject, body, \emph{KeySystemCalls}) and four different document fields (file names, function names, variable names, and Source code with comments).

Thus, our approach obtains twelve different scores
for the source files. Next, we sum the document scores across all twelve searches and calculate the average score, $S_{avg}$ of each file as follows, \[ S_{avg} =  \sum_{n=1}^{12}S_n(q_i,d_j)/12 \] Here, $q$ and $d$ represent each different query and document fields respectively. The value of $i$ is 3 and the value of $j$ is 4. Then we calculate the average score of each file for all the 12 scores. Thus, we get a rank list of the most relevant source files with bug reports.
The benefit of this model is that terms appearing in multiple document fields of the source code are implicitly assigned greater weight since the contribution from each term is summed over all fields in which it appears \cite{structuredIR}.
\subsubsection{Finding Instrumenting Locations}
\label{Approach3a1}

Our goal is to identify system calls to instrument the source code in the potentially buggy files. One challenge is that each system call may occur multiple times in a source file. For example, in the buggy file \emph{copy.c} for the bug report described in Figure~\ref{fig:sample_bugrepo2}, there are nine system calls, each of which appears up to 35 times in that file. Therefore, we must 1) identify the relevant system calls and 2) narrow down their locations in the target source files.
\begin{table*}[t]
\begin{center}
\caption{Steps in generating ranked search list of the system calls form KeySystemCalls}
\label{keysyscall} 
\scalebox{.74}{
\begin{tabular}{|c|c|c|c|c|}
\hline
\emph{KeySystemCalls} & \emph{Tid} &\multicolumn{2}{c|}{\emph{Apriori Step}} & \emph{RankedSearchList} \\
\hline
& &Itemsets(lvl:2)&frq&\\ 
\cline{3-4}
unlink rename link & 1. unlink rename& $\{$unlink$\}$ & 2 & 1. $\{$unlink, rename$\}$ \\
rename rename & 2. link rename rename rename & $\{$rename$\}$ & 5 & 2. $\{$rename, link$\}$ \\

rename unlink &3. unlink rename & $\{$link$\}$ & 1& 3. $\{$rename$\}$\\
rename & & $\{$unlink, rename$\}$ & 2 & 4. $\{$unlink$\}$\\
& & $\{$unlink, link$\}$ & 0 & 5. $\{$link$\}$\\
& & $\{$rename, link$\}$ & 1 & \\
\hline

\end{tabular}
}
\end{center}

\end{table*}

\noindent
\textbf{Identifying relevant system call names.}
\ignore{
This approach first uses the simple keyword search to find the system call 
names in the bug reports. In  Linux system, 
there is a total of 231 system call names and numbers.
The approach matches the bug report against the system call names and identifies
the system calls presented in the bug report as \emph{KeySystemCalls} (
leftmost column of the table \ref{keysyscall}). 
For example, in the bug report of  \emph{mv} program~\cite{mvbug},
the approach identifies 21 system call names.

Our empirical study shows that there 
are 38$\%$ of bug reports where the system call 
names are not stated directly (figure \ref{fig:sample_bugrepo2}). 
In this case, the \emph{KeySystemCalls} is empty. 
To address this problem, we use the manuals of the system calls
to bridge the lexical gap between the system call names and
the bug report. The approach uses 231 different files
containing the system call name along with the description 
under the ``NAME" tag of the Linux manual page. 
The approach then calculates the similarity score of 
the bug report between each of the 231 files. 
Here, the bug report is the query and the 231 documents are the search documents. 
We use the TF.IDF model similarity matrix \cite{TFIDF} to calculate the similarity scores. The output
is a list of ranked system call names in terms
of the scores. The higly ranked system calls
are more relevant to the reported bug. 
\commentty{how many of them will be selected
in the list? top-n?}
The top-n ranked system calls are saved
into a \emph{KeySystemCalls} list. }
To narrow down the locations of the identified system calls, we create a list of system call pairs and rank them. The insight is that buggy interleaving is often related to a system call pair.

We use the \emph{apriori} algorithm~\cite{AprioriAlgo} to identify the system call pairs that exist in the bug report. Apriori is an algorithm for discovering frequent item sets in transaction databases \cite{AprioriAlgo, SPMF}. Using this algorithm, we identify the most frequent system call pairs and system call names in the bug report. Our assumption is that the most frequent system call pair is the most important one and is hence ranked at the top.

To apply the \emph{apriori} algorithm \cite{AprioriAlgo}, we divide the \emph{KeySystemCalls} list based on the \emph{"."} punctuation sign in the bug report from which the system calls are extracted.
The rationale behind this is that in human-written bug reports, if two system calls are related, they are likely to be stated in the same sentence. Our study shows that this happens in 78$\%$ of cases in the bug reports.
Each divided part of the \emph{KeySystemCalls} is considered a transaction for the \emph{apriori} algorithm \cite{AprioriAlgo}. The second column of Table \ref{keysyscall} shows three transactions derived from the \emph{KeySystemCalls} (first column of Table \ref{keysyscall}). Each system call name in the transaction is considered an item (for example, in Tid 1, \emph{unlink} and \emph{rename} are two different items). We apply the \emph{apriori} algorithm on these transactions, setting the itemset (group of items) generation level to two and the minimum support to 100$\%$. The rationale behind this is that we are only interested in system call pairs that appear at least once in the bug report in a sentence. 

In our example, the third column of Table \ref{keysyscall} shows the itemsets containing one and two items. The fourth column of Table \ref{keysyscall} displays the frequencies (number of occurrences) of these itemsets in the bug reports. The fifth column of Table \ref{keysyscall} presents the final ranked list of system call pairs and individual system calls based on their frequencies. A minimum support of 100$\%$ means that if any itemset appears only once in the bug report, we include it in our list. Based on this, we prune an itemset (\emph{unlink, link}) from our list (since it never occurred in any transactions). After pruning, we rank the itemset list based on their frequencies in the bug report. Itemsets containing two items (pairs) are placed at the top of the list, and single items are ranked below the pairs according to their frequencies (column 5 of Table \ref{keysyscall}).

If the bug report does not directly include system call names and the \emph{KeySystemCalls} list contains only derived system call names, we assign a uniform frequency of 1 to all system calls and search for them sequentially in the top-ranked suspicious buggy files. If the \emph{KeySystemCalls} list is empty, we rely on the buggy file list to determine the location. In this case, we instrument each system call one by one in the top-ranked buggy file first, then move to the second-ranked file, continuing this process. However, this approach represents a worst-case scenario, that we did not encounter in our experimental subjects. This highlights the importance of having a high-quality bug report for the success of this automated bug-report-based tool.

\noindent
\textbf{Locating the systems calls in the source code.} 
We use keyword search to find the top-ranked system call pair in the top-ranked source file (Section \ref{rankfile}). In the example provided, the top-ranked system call pair is $\{$unlink, rename$\}$  (first element of column 5 of Table \ref{keysyscall}). We search for this pair in the top-ranked file.
If two system calls of a pair appear in one function one after another in the source code, we rank them according to their order. Otherwise, if the system calls of a pair appear in different functions, we check the function call graph to identify their order.

For example, the \emph{unlink} system call is ranked first because its source file is ranked first, and within that file, it appears before \emph{rename} in the pair. Similarly, the \emph{rename} system call is ranked second because it appears after \emph{unlink} in the pair. In this manner, we search all system calls in our ranked list one by one in the identified top-ranked files. We instrument system calls sequentially. If the first instrumentation can reproduce the bug, then we do not perform any further instrumentation.

If the first instrumentation cannot reproduce the bug, we search for instrumentation points again. We then move to the next file on the list generated by the Structured IR technique, continuing this process until we examine n top-ranked files. Here, we set n to 10, although our experimental results in Section \ref{roleofalgo} show that \Name{} can identify the buggy source files ranked between 1 and 3.

\subsection{Reproducing Bugs}
\label{bugreproduce}
To reproduce system-level concurrency bugs, both input and interleaving information are required.
We have already collected a ranked list of system call locations that are likely to be part of the buggy interleaving. In this step, we will create test cases and perform dynamic instrumentation of the application to reproduce the bug. \Name{} does not aim to replace existing test input generation techniques \cite{10.1145/3238147.3238204, hasan2025llput, DBLP:conf/sigsoft/YuZW17}; instead, it focuses on automating the extraction and enforcement of failure inducing system call interleavings, and can seamlessly integrate with existing input generation tools.

\noindent
\textbf{Creating test cases.}
In order to create test cases, it is necessary to have information about the processes required to reproduce the bug. Our study revealed that $98\%$ of bug reports provide such information. Therefore, in this work, we assume that we already know the application names that produce the bug.

Based on our analysis of system-level concurrency bug reports and previous research \cite{10.1145/3238147.3238204}, we divide the test case of an application into three elements: (i) command (e.g., mv, bash, rm, etc.); (ii) options (e.g., -m, -l, -b); and (iii) input data (e.g., file, folder). To extract the test cases from a bug report, we first perform regular expression matching. If all three elements are present in the bug report, we extract them. If not, we extract the relevant elements and create a set of test cases based on system parameters and our knowledge of functionalities \cite{Causevic10}.

To combine these parameters and environmental conditions into test cases, we use the category-partition method \cite{Ostrand88}, which employs a Test Specification Language (TSL). Test Specification Language (TSL) is a domain-specific language designed to describe test cases, constraints, and expected outcomes in a structured and machine-readable form, enabling automated test generation and execution \cite{Silva2019TSL}. In this approach, parameter choices are selected based on information extracted from bug reports. Similar strategies have been widely adopted in prior research \cite{10.1145/3238147.3238204, Silva2019TSL, Scminer}.  The use of TSL in \Name{} serves only as a structured input representation and does not require specialized expertise. This component is modular and can be replaced by existing automated test input generation techniques.

\noindent
\textbf{Manual Annotation:} To reduce the bais in human gegerated test cases, we follow manual annotation process by two different annotatotors and calculate the aggreement by folowwing formula:
\begin{equation}
    P= \frac{Number of agreements}{Total number of items}
\end{equation}
We find that the agreement value \textit{P} is 0.958,indicating a high level of consistency between annotators

In the future, we plan to enhance this test case generation technique by applying natural language processing (NLP) \cite{nltk} to extract input elements. This could lead to a more accurate and efficient test case generation process without human inervention.

\noindent
\textbf{Dynamic instrumentation.}
To clarify the process further, we first identify the relevant system calls from the bug report as described in the previous steps. Next, we instrument these system calls using PIN~\cite{Pin_A_Dynamic_Binary_Instrumentation_Tool}, an open-source binary instrumentation tool, by inserting a sleep function between each relevant system call pair of the buggy application. If the relevant system call pairs cannot be identified in the source code, we insert the sleep function before and after each relevant system call.

After attaching the instrumentation function, we run the process with the created test case. If the bug is not reproduced in the first run, we move to a new system call and repeat the process by instrumenting it. We instrument the system calls in a ranked manner, starting with the highest-ranked system call and moving downwards. During each run, a specific interleaving of system call execution is exercised with the help of the sleep function, which can cause the bug to reproduce.

By manipulating the order of system call execution for each run, we increase the chances of reproducing the bug. This approach can help us identify the root cause of the bug and assist developers in fixing it.

%% file: experiment.tex
\section{Experiments}
We implement \Name{} as a software tool based on several open-source platforms. We used Gensim \cite{gensim}, a Python library for document indexing and similarity retrieval. We also used  NLTK \cite{nltk}, a platform for building Python programs to work with natural language data. We used NLTK stop words and stemming techniques to pre-process bug reports. We used srcML \cite{srcml} for static analysis. This tool first converts C  code to XML document and then performs queries on it. For applying the apriori algorithm on our dataset we use the SPMF, an open-source data mining library \cite{SPMF}. For instrumenting the source code to reproduce a bug we used PIN \cite{Pin_A_Dynamic_Binary_Instrumentation_Tool}, a dynamic binary instrumentation tool. Our experiments were conducted on a computer with an Intel Core i5-2400 3.10 GHz CPU, 16 GB RAM and Ubuntu 20.04 Linux.

To evaluate \Name{}, we consider four research questions:

\noindent
\textbf{RQ1:} How effective and efficient is \Name{} at localizing buggy interleavings and reproducing the bug?

\noindent
\textbf{RQ2:} What are the roles of the algorithms (structured IR and apriori)  in improving the effectiveness and the efficiency of \Name?

\noindent
\textbf{RQ3:} Does \Name{} benefit developers compared to random reproduction?

\noindent
\textbf{RQ4:} Can \Name{} reproduce bugs from different quality levels of bug reports? 
\label{subsec:benchmarks}

\begin{table*}[t]
\begin{center}
\small
\caption{Benchmark, Ground truths, and Resource information of the Failures}
\label{bench} 
\scalebox{.65}{
\begin{tabular}{|l|r|r|r|l|l|l|}
\hline
Prog. & $NLOC$ & $NOF$ & $NO_Fn$ & \emph{Bug ID} & 
\multicolumn{2}{c|}{\emph{Ground Truths}}\\
\cline{6-7}
& & & & & \emph{File} & \emph{Buggy interleaving loc. (Func., Line No.)}\\
\hline
mv(1) & 7,002 & 321 & 77 & Bugzilla-438076  
& copy.c & copy\_internal;1295,1426 \\
\hline
rm(1) & 5525 & 640 & 76 &  Bugzilla-1211300 
& fts.c & fts\_build,302,fts\_open,490\\
\hline
mkdir(1) & 4,213 & 255 & 26 & Debian-304556 
& mkdir.c & main;161, 189 \\
\hline
mknod(1)& 3,840 & 255 & 26 & Debian-304556 
& mknod.c & main;215,231 \\
\hline
mkfifo(1)& 3,959 & 255 & 26 & Debian-304556 
& mkfifo.c & main;132, 141\\
\hline
ln & 3,890 & 288 & 81 & Debian-357140 
& ln.c & do$\_$link; 287, 310\\
\hline
chmod & 3,983 & 631 & 57 & GNU-11108 
& chmod.c;fts.c & main,263, fts$\_$open,518\\
\hline
pxz & 370 & 1 & 5 & Bugzilla-1182024 
& pxz.c & main,405, 390\\
\hline
gzip(1) & 7,252 & 36 & 35 & Debian-303927 
& gzip.c & treat$\_$file,708;copy$\_$stat,1665\\
\hline
gzip(2) & 6,972 & 34 & 34 & Bugzilla- 155746 
& gzip.c & copy$\_$state, 1627, 1637\\
\hline
bzip2(1) & 9,263 & 12 & 136 & Debian-303300 
& bzip2.c& compressStream,466; compress,1369\\
\hline
bzip2(2) & 9,263 & 12 & 136 & Bugzilla-155744 
& bzip2.c &applySavedMetaInfoToOutputFile,1133,1137\\
\hline
bash & 39,102 & 226 & 456 & Debian-283702 
& histfile.c& history$\_$do$\_$write 436,511\\
\hline
findutils & 32538 & 327 & 271 & Debian-67782 
&copy.c&copy$\_$internal, 1177,1198\\
\hline
locate & 32,538 & 321 & 271 & Debian-461585 
&copy.c&copy$\_$internal, 1177,1198\\
\hline
logrotate & 3,337 & 4 & 45 & Bugzilla-680798 
&logrotate.c & createOutputFile,192,198\\
\hline
coreutils & 12,012 & 255 & 78 & Bugzilla-155758 
& mkdir.c & main, 161, 189\\
\hline
cp & 255972 & 714 & 82 & Debian-649045 
&cp.c& make$\_$dir$\_$parents$\_$private,427,476 \\
\hline
tail & 205113 & 547 & 75 & GNU-548439 
& tail.c& tail$\_$forever$\_$inotify,1290,1366\\
\hline
mv(2) & 7,002 & 321 & 77 & Bugzilla-1141368 
& copy.c & copy\_internal;1295,1426 \\
\hline
mkdir(2) & 4,213 & 255 & 26 & Bugzilla–155760 
& mkdir.c & main;161, 189 \\
\hline
mknod(2) & 3,840 & 255 & 26 & Bugzilla–155760 
& mknod.c & main;215,231 \\
\hline
mkfifo(2)& 3,959 & 255 & 26 & Bugzilla–155760 
& mkfifo.c & main;132, 141\\
\hline
rm(2) & 5525 & 640 & 76 & Bugzilla-1211300 
& fts.c & fts$\_$build,302,fts$\_$open,490\\
\hline
\end{tabular}}
\begin{flushleft}
\small{
$NLOC$ = the number of non-comment lines of code. $NOF$ = the number of C files in the source code. $NO_Fn$ = the number of functions. $File$= the file which contains the buggy interleaving. $Buggy interleaving location$ = Function name, Line number}
\end{flushleft}
\normalsize
\end{center}
\end{table*}

\subsection{Benchmarks}
Our benchmarks consist of real-world Linux application bug reports. These reports describe known concurrency failures caused by incorrectly shared resources between processes and/or signal handlers. We identify these benchmarks by searching open-source repositories such as GNU, Bugzilla, and Debian. We select bug reports that researchers have used in previous studies \cite{Scminer, DBLP:conf/sigsoft/YuZW17,7774517} and confirm as reproducible. Since we need to verify the correctness of our method, we only use benchmarks that previous research has examined and confirmed in this field \cite{Scminer, DBLP:conf/sigsoft/YuZW17,7774517, 10.1145/3476883.3520207}.

However, we do not include all 21 benchmark applications from Descry \cite{DBLP:conf/sigsoft/YuZW17} in our study. This limitation exists because three application bugs in that work lack associated bug reports and originate instead from developer changelogs. In this study, we use 24 bug reports from 19 program versions across 17 unique applications as benchmarks.
We only know that the bugs are reproducible; the root causes of these failures remain unknown until we complete the execution and analysis of \Name{}. Descry \cite{DBLP:conf/sigsoft/YuZW17}, a related tool, conducts its evaluation in a similar manner. After analyzing the results produced by \Name{}, we manually examine the bug report comments, patches, developer discussions, and earlier research works \cite{Scminer, DBLP:conf/sigsoft/YuZW17,7774517, 10.1145/3476883.3520207} in this field to determine the ground truth and validate the accuracy of \Name's output.
In this paper, we examine the highest number of benchmarks compared to prior works. Descry \cite{DBLP:conf/sigsoft/YuZW17} includes 21 benchmarks, Scminer \cite{Scminer} includes 19, and RRF \cite{7774517} analyzes 7 benchmarks. Therefore, the total number of benchmarks in our experiment remains comparable to previous studies.

Before applying \Name{} on a bug report, we only knew that the bug was reproducible. We did not know the root causes or the locations of the failures until we finished running and analyzing the results of \Name{}. Table \ref{bench} shows the statistics of the benchmarks. For each benchmark, we record its number of non-comment lines of code (NLOC), number of source files (NOF), number of functions ($NO_{fn}$), bug report ID, and ground truths. The ground truths specify the file name where the buggy system call resides (Column 6 of Table \ref{bench}). The last column of the table presents the source code location, including the function name and line number, where we need to instrument the code to reproduce the bug. Each bug is associated with two system calls, and the line numbers indicate where these system calls appear. 


\subsection{Evaluation Metrics}
\textit{Localize buggy interleavings.} 
\Name{} computes the top-$N$ system calls where the buggy interleavings can be found. We set $N$ to 20 because our experimental results show that, in successful scenarios, at least one system call from the buggy interleaving is ranked in the top 20 positions (Column 7 of Table \ref{results}). To assess the effectiveness of localizing buggy interleavings, we measure three metrics. The first metric measures the rank number (position) of system call names identified by \Name{}. The ground truths are determined by manually examining the solution discussed in the corresponding bug report, the patch used for fixing the issue, and then by reproducing the bug.

For the second metric, we use Mean Average Precision (MAP). MAP is a single-figure measure of ranked retrieval results independent of the size of the top list \cite{MAP}. It is designed for general ranked retrieval problems, where a query can have multiple relevant documents (e.g., a buggy interleaving may be associated with more than one system call in different places of the source code).

To compute MAP, we first calculate the average precision (AP) for each individual query $Q_i$, and then calculate the mean of APs over the set of queries:

\begin{center}
$\mathit{MAP = \frac{1}{|Q|} \cdot \displaystyle\sum_{Q_i \in Q} AP(Q_i)}$
\end{center}

To illustrate the MAP calculation, suppose there are bug-related system calls $SC_1$ and $SC_2$. If Technique-I ranks the two system calls at the 1st and 2nd positions among all 500 functions, and Technique-II ranks the two functions at the 1st and 3rd positions, then the MAP of Technique-I is (1/1 + 2/2) / 2 = 1, and the MAP of Technique-II is (1/1 + 2/3) / 2 = 0.8 \cite{Scminer} \cite{6100061}.

For the third metric, we use recall rate@$K$. By fixing the size of the top list to $K$, recall rate@$K$ \cite{introtoIR} measures the fraction of buggy interleavings that are successfully detected in the retrieved top-$K$ locations of the source code ($N_{detected}$), among all the buggy interleavings ($N_{total}$) addressed in the bug report \cite{6100061}.

\begin{center}
$\mathit{recall\ rate@k = \frac{N_{detected}}{N_{total}}}$
\end{center}

\textit{Reproducing the bug from the bug report.} 
We measure both effectiveness and efficiency. To measure effectiveness, we check whether a known failure can be successfully reproduced within 100 attempts. To measure the efficiency of \Name{}, we consider the number of attempts needed to reproduce the bug and analyze the failure reproduction time by measuring the time spent on different steps of our procedure.
\input{results}

%% file: results.tex
\ignore{
\begin{table*}[htb]
\begin{center}
\caption{Results of applying \Name{} over benchmark applications}
\setlength{\tabcolsep}{1pt}
\label{results2} 
\scalebox{0.58}{
\begin{tabular}{|l|c|c|c|c|c|c|c|c|c|c|c|c|c|c|c|c|c|c|c|}
\hline
\emph{Prog} & \emph{mv} & \emph{rm} &  \emph{mkdir} & \emph{mknod} & \emph{mkfifo} & \emph{ln} & \emph{chmod} & \emph{pxz} & \emph{gzip(1)} & \emph{gzip(2)} & \emph{bzip2(1)} & \emph{bzip2(2)} & \emph{bash} & \emph{findutils} & \emph{locate} & \emph{logrotate} & \emph{coreutils} & \emph{cp} & \emph{tail}\\ 
\hline
\emph{Suc.}& Y & Y & Y & Y & Y & Y & Y & Y & Y & Y & Y & Y & Y & Y & Y & Y & Y & Y & Y\\ 
\hline
\emph{NoR} & 1 & 1 & 1 & 1 & 1 & 1 & 2 & 1 & 3 & 3 & 1 & 1 & 1 & N & 1 & 2 &1 & 14 & 1\\
\hline
\emph{T(m)}& 8.15 & 12.31 & 6.21 & 6.21 & 6.21 & 6.87& 23.22 & 1.83 & 4.56 & 6.12 &0.64 &0.75 & 5.41& $>$ & 10.23 & 3.2 & 6.88 & 59.09 & 7.19\\
\hline
\end{tabular}}
\begin{flushleft}
\small
\emph{NoR} = the number of runs to reproduce the bug.
\emph{Suc.} = indicates whether \Name{} is successful to reproduce the bug or not.
\emph{T(m)} = the time spent on the analysis in minutes. 
\end{flushleft}
\normalsize
\end{center}
\end{table*}
}
\subsection{Results and Analysis}
\label{subsec:results}
Table \ref{results} shows the results of applying \Name{} on our benchmark applications.
\begin{table*}[!htb]
\begin{center}
\caption{Results of applying \Name{} over benchmark applications}
\setlength{\tabcolsep}{.72pt}
\label{results} 
\scalebox{0.7}{
\begin{tabular}{|l|c|c|c|c|c|c|c|c|c|c|c|c|}
\hline
\emph{Prog} &\multicolumn{3}{c|}{\emph{Localizing Files}} &\multicolumn{5}{c|}{\emph{Localizing System Calls.}} &\multicolumn{2}{c|}{\emph{Bug Repro.}}& \emph{Time} &\emph{Type}\\
\cline{2-11}
&\emph{BRk} & \emph{ESRk} & \emph{SRk} & \emph{Syscalls} &\emph{ORnk} & \emph{Rank} & \emph{Rec.} &\emph{MAP} & \emph{Suc.} & \emph{NoR} &  \emph{(min)} &\\ 
\hline
mv(1) & 1 & 1 & 1 & unlink,rename & 19,23 & 1,2 & 1 & 1 & Y & 1 & 8.15 & C\\
\hline
rm(1) & 1 & 1 & 1 & fstatat,openat & 2,11 & 1,11 & 1 & .6 & Y & 1 & 12.31 & C\\
\hline
mkdir(1) & 2 & 1 & 1 & mkdir,chmod &1,2 & 1,2 & 1 & 1 & Y & 1 & 6.21 & C\\
\hline
mknod(1) & 8 & 5 & 2 & mknod,chmod &1,3 & 1,2 & 1 & 1 & Y & 1 & 6.21 & C\\
\hline
mkfifo(1)& 4 & 3 & 1 & mknod,chmod &1,2 & 1,2&1 & 1 & Y & 1 & 6.21 & C\\
\hline
ln & 2 & 2 & 2 & stat,unlink & 8,11 & 1,2 & 1 & 1 & Y & 1 & 6.87 & C\\
\hline
chmod & 1, 2 & 1,2 & 1,2 & fchmodat &4,20 & 1,20 & .5 & .55 & Y & 2 & 23.22 & S\\
\hline
pxz & 1 & 1 & 1 & chmod & 45,38 & 1, 33 & .5 & .5 & Y & 1 & 1.83 & S\\
\hline
gzip(1) & 2 & 3 & 2 & chmod & 15,41 & 2,16 & .5 & .3 & Y & 3 & 4.56 & I\\
\hline
gzip(2) &  2 & 3 & 1 & chmod & 3, 18 & 2,18 & .5 & .3 & Y & 3 & 6.12 & I\\
\hline
bzip2(1) & 2 & 1 & 1 &chmod& 12,38 & 1,5 & .5 & .7 & Y & 1 &0.64 & I\\
\hline
bzip2(2)& 3 & 1 & 1 &chmod& 33, 34 & 1,2 & .5 & 1 & Y & 1 &0.75 & I\\
\hline
bash & 7 & 6 & 3 &write& 19,28& 4,6&.5 & .3 & Y & 1 & 5.41 & S\\
\hline
findutils & 0 & 0 & 0 &$*$& $*$ & $*$ & 0 & 0 & N & 100+ & $>$ & I\\
\hline
locate & 0 & 1 & 1 &rename,chmod& 19,23 & 1, 11 & .5 & 0.6 & Y & 1 & 10.23 & S\\
\hline
logrotate & 1 & 1 & 1 & open & 1,9 & 1,2 & .5 & 1 & Y & 2 & 3.2 & I\\
\hline
coreutils & 2 & 6 & 1 &mkdir,chmod& 1,2 & 1,2 & 1 & 1 & Y & 1 & 6.88 & I\\
\hline
cp & 21 & 17 & 1 &chmod& 6,8 & 7,9 & 0 & .18 & Y & 14 & 59.09 & S\\
\hline  
tail & 9 & 12 &2& read, write &4,7 & 1,4& 1 & .75 & Y & 1 & 7.19 & C\\
\hline
mv(2) & 1 & 1 & 1 & unlink,rename & 19,23 & 1,2 & 1 & 1 & Y & 1 & 7.35 & C\\
\hline
mkdir(2) & 2 & 1 & 1 & mkdir,chmod &1,2 & 1,2 & 1 & 1 & Y & 1 & 6.21 & S\\
\hline
mknod(2) & 8 & 5 & 2 & mknod,chmod & 1,3 & 1,2 & 1 & 1 & Y & 1 & 6.21 & S\\
\hline
mkfifo(2)& 4 & 3 & 1 & mknod,chmod & 1,2 & 1,2& 1 & 1 & Y & 1 & 6.21 & S\\
\hline
rm(2) & 1 & 1 & 1 & fstatat,openat & 2,11 & 1,11 & 1 & .6 & Y & 1 & 11.25 & C\\
\hline
\end{tabular}}
\begin{flushleft}
\small
\emph{BRk}  = the ranking position of the ground truth file by applying the basic IR technique. 
\emph{ESRk} = the ranking position of the ground truth file by applying the traditional structured IR technique.
\emph{SRk} = the ranking position of the ground truth file by applying our structured IR technique.
\emph{Syscalls} = System calls of the buggy interleaving. 
\emph{ORnk} = the ranking position of the ground truth without applying the apriori algorithm.
\emph{Rank}= the ranking position of the ground truth by applying apriori algorithm.
\emph{Rec.} = the Recall@K score.
\emph{MAP} = the Map score.
\emph{Suc.} = indicates whether \Name{} is successful to reproduce the bug or not.
\emph{NoR} = the number of runs to reproduce the bug.
\emph{Time} = the time spent on the analysis. 
\emph{Type} = The bug report classification done by manual observation, C= complete, S= semi-complete, and I= incomplete
\end{flushleft}
\normalsize
\end{center}
\end{table*}

\subsubsection{RQ1: Effectiveness and Efficiency of \Name{}}
\Name{} succeeded in reproducing 23 bugs from the bug reports among 24 benchmarks, achieving a success rate of 95.8$\%$. Column 10 titled as \emph{Suc.} of Table \ref{results} marks Y for the bugs that are reproducible by \Name{}. These results indicate that \Name{} is effective in reproducing system-level concurrency failures.

The only bug not reproduced by \Name{} is the \emph{findutils} bug report. The reason is that the report does not contain any direct or indirect information about system call names and lacks the correct application name.

Our results indicate that, for 22 benchmarks, \Name{} can identify at least one system call from the system call pair causing the failure. Column 9, named \emph{Rec.} in Table \ref{results}, shows that the recall@K score ranges from 0 to 1, with an average of 0.75. Here, a value of 0 means \Name{} cannot identify any of the system call names from the buggy interleaving, while a value of 1 means it identifies all the system call names of the buggy interleaving. 

The \emph{Rank} column shows the ranks of the system calls in the buggy interleavings among all the system calls of the application. This column contains multiple ranks because there are multiple ground truths. The MAP score ranges from 0 to 1, with an average of 0.77, indicating that all system calls identified by \Name{} are ranked in the top 5.

The number of iterations to reproduce each bug ranged from 1 to 14, with an average of 2 presented in the second last column of table \ref{results}. The number of iterations depends on the MAP score and Recall@K score. For example, in the pxz benchmark (row 8 of Table \ref{results}), the bug report identifies one system call from the buggy interleaving — chmod, which \Name{} ranks as 1. Another system call from the same buggy interleaving ranks at 33. Since this bug report does not mention any system call pairs, \Name{} cannot identify any pairs. To reproduce the bug, \Name{} instruments the application just before the rank 1 system call (chmod) and runs the program. The bug reproduces on the first attempt. Although \Name{} cannot identify the complete ground truth (resulting in a Recall@K score of 0.5 for this benchmark), it still reproduces the bug after one execution. This works because another system call from the buggy interleaving occurs before chmod. By instrumenting the code between these two system calls, even though only chmod is known, \Name{} successfully triggers the bug.

The process differs for the chmod benchmark (row 7). Here, \Name{} requires two iterations. In the first run, it instruments the code before the identified system call fchmodat (column 5 of Table \ref{results}), but the bug does not reproduce. On the second attempt, it instruments the code after this system call, successfully triggering the bug. This happens because the other system call from the buggy interleaving occurs after the identified system call. We create simple scripts to automate both the instrumentation adjustments and the execution of the buggy applications.

The time required to reproduce the bugs ranged from 0.64 to 59.09 minutes, with an average time of 8.9 minutes. The last column of Table \ref{results} shows the total time to reproduce a bug from a given bug report. These results suggest that \Name{} is efficient in reproducing system-level concurrency bugs from bug reports and is suitable for real-world use.

Our experimental results show that in nine cases (rows 7-13 and rows 15-16 of Table \ref{results}), we identify only one system call from the buggy interleaving. This happens because the corresponding bug reports mention only a single system call. To determine the root cause and locate the buggy interleaving, we need to capture all system calls involved. Therefore, in our analysis, if \Name{} fails to identify all system calls, we mark the \emph{Rec.} column (column 8 of Table \ref{results}) accordingly. Even when we identify just one system call, we can still reproduce the bug in most cases within two iterations. First, we place the instrumentation point before the identified system call and run the program. If the bug reproduces successfully, we stop there. This approach works in 8 out of the 9 cases where the bug report mentions only one system call. For the \emph{logrotate}, we run the program twice. In the first attempt, we place the instrumentation point before the system call, but the bug does not reproduce. In the second attempt, we move the instrumentation point after the system call, which successfully reproduces the bug.

In this paper, we utilize the Apriori algorithm to identify and rank both system call pairs and individual system calls. When system call pairs cannot be extracted from bug reports, we instead extract single system call names. Identifying faulty system call pairs ensures reproducing the bug in a single iteration.  
\subsubsection{RQ2: Roles of the Algorithms} 
\label{roleofalgo}
\textbf{Sturctured IR for \Name{}.} Columns 2 (BRnk), 3 (ESRnk), and 4 (SRnk) of Table \ref{results} show the rank of the file containing the buggy interleaving using the basic IR technique, the traditional structured IR technique, and our structured IR technique, respectively. These values indicate that, by using the structured IR technique, we get our necessary files ranked from 1 to 3. We failed to identify the respective file in \emph{findutils} because the bug report does not contain the application's name.

\begin{figure}[htb]
\centering
\includegraphics[scale=0.8]{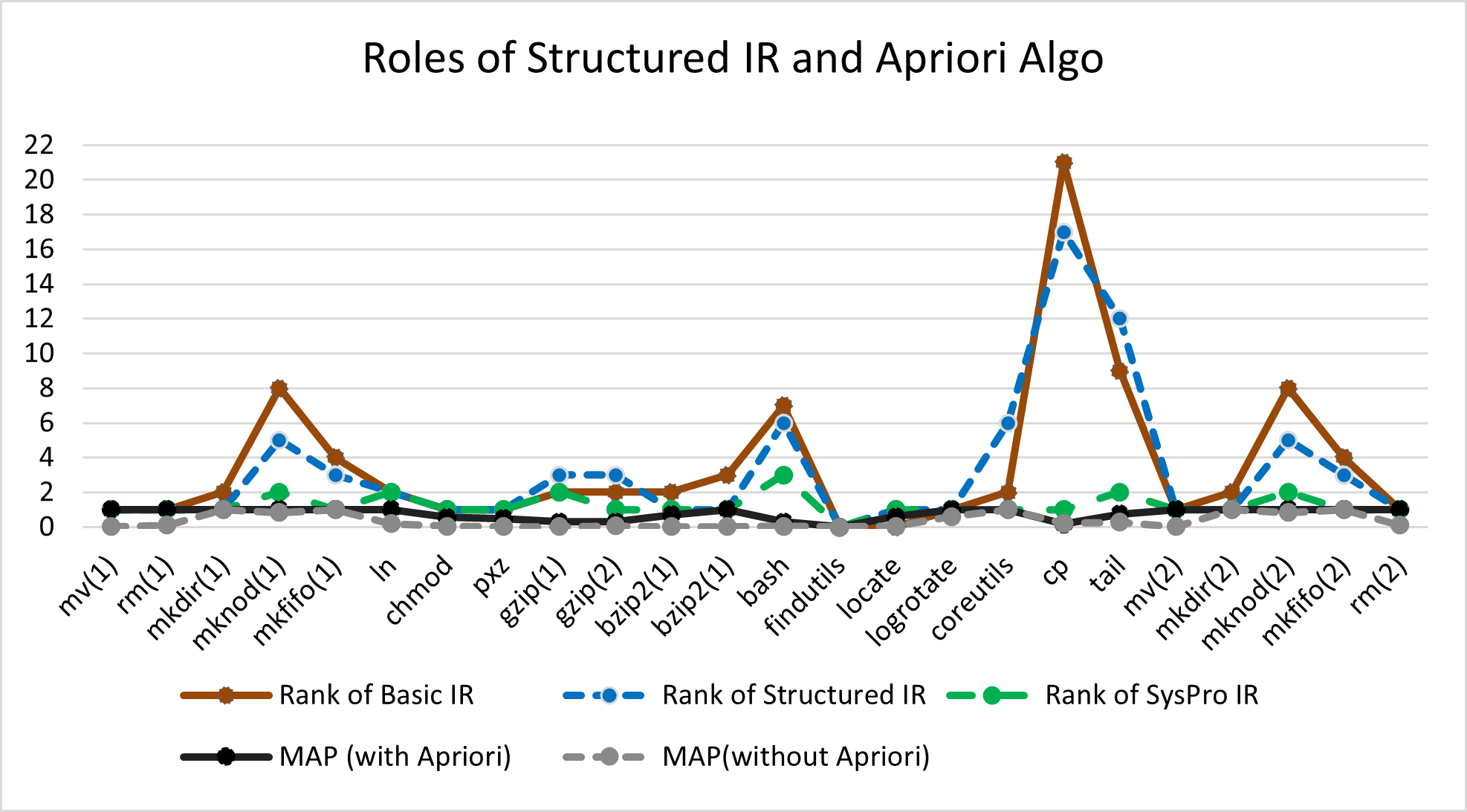}
\caption{Roles of Structured IR and Apriori Algorithm in improving the effectiveness and the efficiency of \Name{}}
\label{fig:rq2}
\end{figure}

When comparing the file ranks, we find that, in 14 cases, the our proposed structured IR technique performs better than the basic IR technique, and in 10 cases it performs better than the traditional Structured IR technique\cite{structuredIR}, which does not distinguish queries for system calls. Figure \ref{fig:rq2} shows the comparison between these three techniques. On average, the ranking is improved by using our structured IR technique compared to the basic IR technique, with improvements ranging from 0 to 100$\%$, and an average improvement of 54.5$\%$. Compared to the traditional structured IR technique, the improvements range from 0 to 100$\%$, with an average improvement of 63.8$\%$.

\textbf{Apriori for \Name{}.} Column 6 (ORnk) and Column 7 (Rank) of table \ref{results} show the ranking of the system calls before and after applying the data mining algorithm, respectively. The ranking of the system calls significantly increases after 
applying the algorithm for 17 subjects, remains unchanged for 6 subjects, and decreases for one subject (\emph{cp}). It decreases in the case of \emph{cp} since the bug report of \emph{cp} contains a different system call name, which is not in the buggy interleaving. For this reason, our technique gives that system call top rank. Figure \ref{fig:rq2} shows the comparison between the MAP scores i) by using the apriori algorithm and ii) by not using that. 
The comparison shows that on average, the algorithm improves 
the MAP score from 0 to 80$\%$ with an average of 46$\%$.

\begin{figure}[htb]
\centering
\includegraphics[scale=0.8]{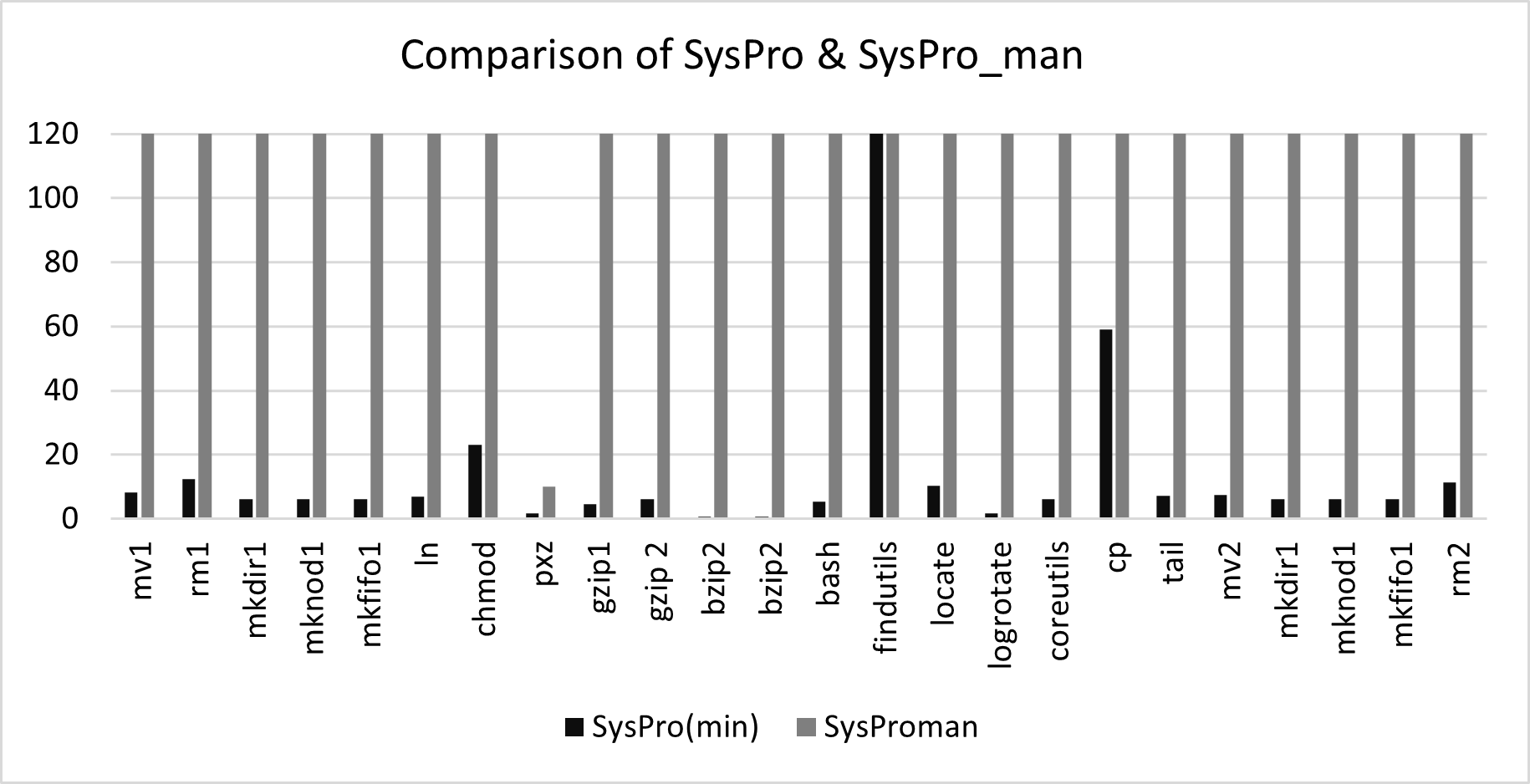}
\caption{\Name{} benefits developers compared to random testing}
\label{fig:rq3}
\end{figure}

We apply the Apriori algorithm to identify and rank system call pairs, which reduces the number of iterations needed to reproduce bugs. For example, in the \emph{mv(1)} bug report, extracting a single system call like unlink shows that it appears four times in the source code. This process may require up to 4×2=8 iterations to reproduce the bug. In contrast, our approach identifies the exact system call pair, allowing us to reproduce the bug in a single iteration. Since each iteration increases the total bug reproduction time, we prioritize extracting system call pairs from bug reports when they are mentioned. 

\subsubsection{RQ3: Comparing with Random Testing} 
To the best of our knowledge, no existing tools can automatically reproduce system-level concurrency bugs from bug reports. We compared the total running time of \Name{} with random testing, referred to as \Name$\_$man. \Name$\_$man uses the same input as \Name{} but without instrumentation. Previous research \cite{7774517} also compares their technique with this kind of random stress testing. In this type of random testing, we run the buggy processes together randomly multiple times to reproduce the bug.

If a bug can be reproduced by \Name$\_$man within a 120-minute (2-hour) time limit, we mark it as a success and record the time. If it cannot be reproduced within the time limit, we mark it as a failure. We set the time limit to 2 hours because previous research \cite{DBLP:conf/sigsoft/YuZW17} uses this time limit for random testing. Within this 2-hour time limit, we run the buggy processes together as many times as possible.

In Figure \ref{fig:rq3}, when a bar reaches the top of the vertical axis, it means the corresponding method failed to reproduce the failure within the time limit. \Name{} succeeded in 18 benchmarks, whereas \Name$\_$man failed in all 19 cases.
\begin{figure}[htb]
\centering
\includegraphics[scale=0.8]{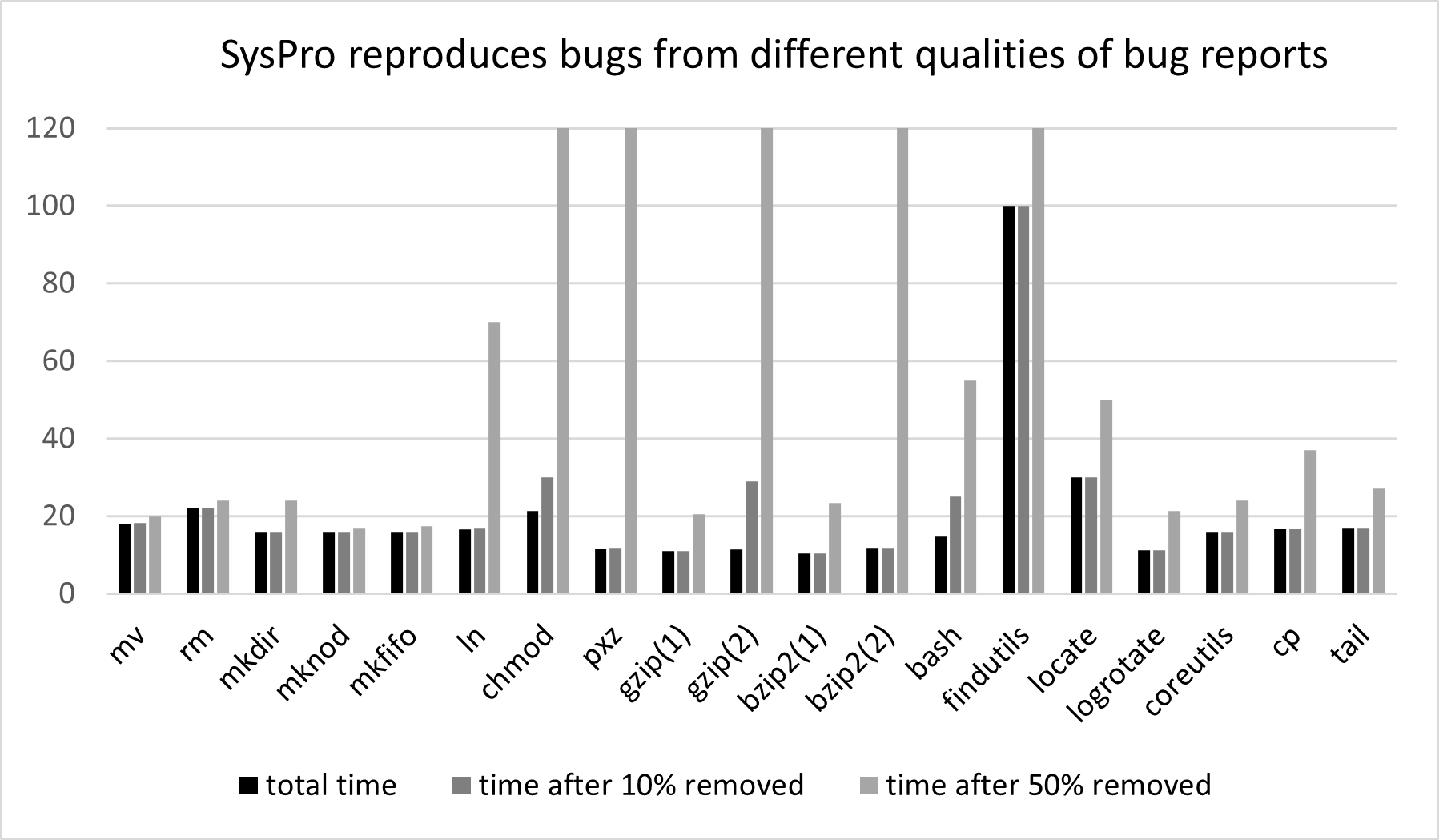}
\caption{ \Name{} can reproduce bugs from different qualities of bug reports}
\label{fig:rq4}
\end{figure}

\subsubsection{RQ4: Handling Different Qualities of Bug Reports.} 
In our study, we define a classification scheme for bug reports to better assess their quality and evaluate the performance of our technique, \Name{}. A bug report is categorized as complete if it explicitly contains both the input and interleavings required to reproduce the failure. Reports that provide only one of these components, either input or interleaving information directly, are classified as semi-complete. All remaining reports, that convey the information indirectly, are labeled as incomplete. It is important to note that this classification is introduced solely for the purposes of our study; it is not an established taxonomy and is unrelated to conventional bug statuses such as ``fixed" or ``closed." This categorization enables a more structured analysis of how \Name{} performs across bug reports of varying completeness levels. This classification is derived through manual observation, and no established ground truth for it currently exists.

The last column of Table \ref{results} presents our manual classification for each bug report. The results indicate that \Name{} successfully reproduces all bugs from both the complete (9 in total) and semi-complete (8 in total) bug reports, but it failed to reproduce one bug from an incomplete (7 in total) bug report. The success rate for the incomplete bug reports is 86\%. This failure is attributed to the absence of the buggy application's name in the report. These experimental results demonstrate that \Name{} can reproduce bugs from bug reports where the relevant information is either directly or indirectly provided.

To address \emph{RQ4}, We randomly remove 10\% and 50\% of the words from our benchmark bug reports and compare the results. This removal process does not isolate system call names or any specific structure — it is entirely random. As a result, system call names may be removed in some cases and remain intact in others. Figure \ref{fig:rq4} shows a chart where we compare the times of reproducing bugs from: i) the original bug report, ii) the bug report with 10$\%$ of words removed, and iii) the bug report with 50$\%$ of words removed. The experiment shows that our technique succeeded in all cases when 10$\%$ of words were removed from the bug report. However, in some cases, the total execution time increased. This is because if any system call name is removed from the bug report, its rank decreases, requiring more iterations to reproduce the bug.

After removing 50$\%$ of words from the bug reports, \Name{} failed to reproduce the bug for 4 subjects within the given 120-minute time limit. This mostly happened due to the size of the bug reports. Some bug reports contain just the necessary information to reproduce the bug, so removing 50$\%$ of the content resulted in the loss of all essential information. Consequently, more time than our threshold was needed to reproduce those bugs. Word removal doesn’t inherently lead to low- or high-quality bug reports. If essential information, such as system call names, function names, file names, or code snippets gets removed, the report's quality decreases. However, if unnecessary text is removed while keeping essential details intact, the report’s quality may improve. For this reason, we explore bug reports with randomly altered quality rather than categorizing them as high or low quality. The results presented here reflect this random variation in bug report quality.
\subsection{Compare with Existing Works.} 
\textbf{\Name{} Vs. Descry:} We cannot directly compare \Name{} with the recent work Descry \cite{DBLP:conf/sigsoft/YuZW17}, which also aims to reproduce system-level concurrent bugs. This distinction arises because Descry relies on console logs to reproduce a bug, and not all bug reports include such logs. Specifically, the bug reports used in our experiment do not contain separate console logs. In contrast, \Name{} utilizes only the bug report's subject, summary, and description to reproduce the bug. This approach accommodates real-world scenarios where we may only have a bug report available for reproducing and analyzing the issue. While both techniques are designed to address system-level concurrency bugs, they differ in the resources they require for analysis. Furthermore, it is important to note that there are currently no other fully automated techniques available to reproduce this particular type of bug.

\textbf{\Name{} Vs. RrF:} \Name{} automatically identifies the relevant system call name, or instrumentation location, from the bug report without human intervention. In contrast, RrF \cite{7774517} relies on engineers to manually analyze the bug report and extract the necessary information for reproducing the bug. Since no established metric evaluates human expertise in extracting information from bug reports, comparing RrF’s manual approach to \Name{}'s automated technique is not meaningful. Additionally, RrF requires application configurations, two processes undergoing debugging, test inputs, bug reports, and failed outputs. However, we do not have all these components for every subject; we only have bug reports.

%% file: discussion.tex
\section{Discussion and Limitations}
Many related works \cite{7774517, Simracer, 10.1145/2365864.2151052, 10.1145/3476883.3520207} assume that inputs are known to developers, and these approaches have demonstrated effectiveness in real-world settings. Several existing tools also focus on test case generation \cite{DBLP:conf/sigsoft/YuZW17, 10.1145/3238147.3238204, hasan2025llput}. Such tools could be integrated with \Name{} to generate inputs in future work. \Name{} is orthogonal to input generation tools \cite{DBLP:conf/sigsoft/YuZW17, 10.1145/3238147.3238204}. In practice, test cases are often available to developers; among them, developers must identify the failure-inducing ones, and bug reports frequently provide hints about the relevant inputs. Therefore, we assume that manually generating test cases is neither uncommon nor particularly difficult for developers. However, the central challenge in system-level concurrency bugs lies in identifying the system call pairs that trigger the bug. None of the existing work addresses this challenge by automatically identifying such system call pairs. In contrast, our approach fully automates the identification of the system call pairs responsible for triggering these bugs. Although reliance on partial manual input generation is a limitation, it does not affect the validity of interleaving identification.

\subsection{Bug Causing Process. } \Name{} primarily focuses on extracting system calls related to buggy interleavings for the specified application detailed in the bug report. However, reproducing a concurrency failure often demands both specific inputs and interleavings. We assume that users know the set of processes contributing to the concurrency bugs. Although the problematic process is typically indicated (approximately 98$\%$) in the bug report, information about other processes is not consistently included. Consequently, extracting additional process information from the bug report is not always feasible, necessitating prior knowledge about the processes. The effectiveness of \Name{} is heavily reliant on the quality of the provided bug reports.
\subsection{Future Scopes.} As \Name{} is the first approach to reproduce system level bugs by using the bug reports, there are many existing challenges researchers can address and provide also provide a variety of future research scopes.\\
\textbf{Benchmark Collection: } The success of \Name{} depends on the quality of bug reports. In this paper, we analyze 24 bug reports. Collecting benchmarks for system-level concurrency bugs presents significant challenges. First, None of the bug reports are explicitly classified as system-level concurrency bug reports. Since the same application can contain both regular serial bugs and system-level concurrency bugs, distinguishing between these types of bug reports is challenging. 

Second, existing research confirms only 21 reproducible system-level concurrency bugs.

Third, experts must first verify these bugs before any reproduction attempts can be made. However, finding experts with the specialized knowledge to assess this type of bug is difficult. Fourth, Since manually reproducing them is difficult and no automated techniques currently exist, we develop \Name{} to address this challenge. Researchers can use \Name{} to reproduce system-level concurrency bugs from bug reports and contribute to a dataset of reproducible cases, benefiting the research community. Our goal is to expand this dataset, creating a more comprehensive resource for system-level concurrency bug research.\\
\textbf{Extracting Inputs:} In this paper, we focus on the challenges of extracting instrumentation locations. We use the TSL method to generate inputs; however, this approach is relatively naive and may not efficiently extract inputs from bug reports. In the future, we plan to explore more advanced methods, such as recent LLM techniques, to extract input-related information more effectively. Furthermore, while our current approach targets system-level bugs, we aim to extend its applicability to concurrent bugs and web-based sequential bugs, evaluating its performance across these categories.\\
\textbf{Usability Check:} We plan to develop a user-friendly tool based on the \Name{} framework. Researchers can explore the tool and conduct usability tests. Based on their findings, they can propose more user-friendly techniques for debugging and reproducing concurrency bugs.
\section{Threats to Validity}
The primary threat to the external validity of this study involves the representativeness of our subjects and bug reports. Other subjects may exhibit different behaviors. Data recorded in bug tracking systems and code version histories can have a systematic bias relative to the full population of bug reports and can be incomplete or incorrect. However, we mitigate this threat to some extent by using several varieties of well-studied open-source code subjects and bug sources for our study \cite{Scminer,DBLP:conf/sigsoft/YuZW17,7774517}. We cannot claim that our results can be generalized to all systems in all domains.

The primary threat to internal validity arises from the use of keyword searches and manual inspection to identify subject bug reports. If a bug report omits relevant system calls and associated keywords, the proposed technique may fail to identify it. Consequently, the effectiveness of \Name{} heavily depends on the quality and completeness of the provided bug reports. Nevertheless, the combination of keyword searches and manual inspection has proven effective for identifying specific types of bugs from large collections of generic bug reports and has been successfully applied in prior studies \cite{Scminer, DBLP:conf/sigsoft/YuZW17, 10.1145/3238147.3238204, 10.1145/3476883.3520207}. 

The primary threat to construct validity involves the dataset and metrics used in the study. To mitigate this threat, we used bug reports from the bug tracking systems of the 19 subjects, which are publicly available and generally well-understood. We also used well-known, accepted, and validated measures of ranking scores, MAP, and recall rate@$K$.

We minimized threats to conclusion validity by performing the same experiment on the same bug reports repeatedly, maintaining the same setup and input throughout.

%% file: relatedwork.tex
\section{Related Works}
\noindent
\textbf{Reproducing Bug from bug report.} There has been prior work on automatically reproducing bugs in Android applications using bug reports, leveraging different sources such as textual reproduction steps \cite{10.1145/3213846.3213869,10.1145/3597503.3608137,10172656,10411964,10.1145/3650212.3680341,10.1145/3597926.3598066,recdroid,7965246}, video recordings \cite{10.1145/3377811.3380328,10.1145/3510003.3510048}, and crash stack traces \cite{6926857,10.1145/3597503.3623298,10599338}. In particular, tools such as ReCDroid \cite{recdroid}, ReproBot \cite{10.1145/3597926.3598066}, and Roam \cite{10.1145/3660824} aim to automatically reproduce Android crashes from bug reports. Feng et al. \cite{10.1145/3597503.3608137} explored the use of large language models (LLMs) to execute reproduction steps in mobile app bug reports. CrashScope \cite{7965246} is another practical system that automatically discovers, reports, and reproduces crashes in Android applications.

However, unlike Android application bugs, where user interactions and UI event sequences are central to reproduction, system-level concurrency bugs require analyzing system calls, which are a key and distinct feature of such bugs. Therefore, techniques developed for Android bug reproduction are not directly applicable to system-level concurrency bug reproduction.

\noindent
\textbf{Reproducing System-level concurrency bugs.} System-level concurrency bugs, due to their inherently non-deterministic nature, have attracted researchers to work in this field. Various approaches\cite{7774517, DBLP:conf/sigsoft/YuZW17, 10.1145/3476883.3520207, 10.1145/2491956.2462162} have been proposed to automatically reproduce such bugs. However, none of these prior research works focus exclusively on using a bug report as the primary input for bug reproduction. For instance, Yu et al. introduced a fully automated tool\cite{DBLP:conf/sigsoft/YuZW17} that reproduces system-level concurrency failures based on default log messages collected from production field data. Another approach, RRF\cite{7774517}, is mostly automated but requires human intervention to gather information from bug reports or bug descriptions. Reproducing bugs with RRF heavily relies on the knowledge of the engineer analyzing the bug reports, as it necessitates configuration files and information about the processes causing the bug. Another research work, CONCURRIT\cite{10.1145/2491956.2462162} introduced a domain-specific language (DSL) for reproducing concurrency bugs. In this technique, programmers need partial knowledge of the bug and write a test script in the CONCURRIT language to formally and concisely express the bug. This approach also relies on the programmer's knowledge. Additionally, Zaman et al. developed RedPro\cite{10.1145/3476883.3520207}, which detects system-level race conditions through runtime monitoring and then regenerates the race condition by instrumenting the code. Our proposed technique stands apart from these approaches. In our method, we exclusively use bug reports to reproduce the bug, and our approach is entirely automated. It identifies bug-related system call names in the source code and manages their execution order to reproduce the bug.

\noindent
\textbf{Reproducing Crashes Automatically.} Several existing works \cite{10.1007/s10664-019-09762-1}, \cite{6926857}, and \cite{7081820} have focused on reproducing software crashes. For instance, Soltani et al. \cite{10.1007/s10664-019-09762-1} proposed a technique that creates a benchmark of 200 real-world Java crashes and reproduces them using EvoCrash \cite{8502801}, a state-of-the-art tool for search-based crash reproduction. STAR \cite{6926857} presents an automatic crash reproduction framework that utilizes collected crash stack traces. It integrates backward symbolic execution with a novel method sequence composition approach to generate unit test cases that accurately reproduce original crashes without additional runtime overhead. JCHARMING\cite{7081820} employs crash traces and model checking to identify program statements necessary for crash reproduction. By leveraging model checking for completeness while filtering out unnecessary system states using crash trace information and static slicing, JCHARMING enhances crash reproduction accuracy. However, none of these techniques specifically target system-level concurrency bugs and, therefore, cannot be used to reproduce such issues, as they do not consider system call information or test cases from different applications. Additionally, some of these techniques \cite{6926857, 7081820} rely solely on stack traces and are designed only for reproducing Java crashes. To address these limitations, we propose \Name{}, a novel approach for reproducing system-level concurrency bugs. Unlike existing methods, \Name{} focuses on applications written in C/C++ and leverages only the information available in bug reports, making it more applicable for reproducing concurrency-related failures at the system level.

\noindent \textbf{Reproducing Different Kinds of bugs Automatically.} ReproLite \cite{10.1145/2670979.2671004} is designed to reproduce bugs in the Cassandra and HBase cloud systems. It introduces a Domain-Specific Language (DSL) that enables testers to define potential buggy scenarios, including specific workloads, execution sequences, and environmental non-determinism that may trigger a given bug. ReproLite enforces these conditions during execution and can also generate potential buggy scenarios automatically from log sequences identified as indicative of a bug's root cause. While ReproLite specifically targets cloud system bugs, our proposed technique, \Name{}, focuses on reproducing system-level bugs. Unlike ReproLite, \Name{} does not require a DSL; instead, it relies solely on existing unstructured bug reports to automatically reproduce bugs.

Several existing studies have explored automated bug reproduction techniques for Android applications by leveraging different types of resources. Zhao et al. \cite{10.1007/978-3-030-22888-0_8} proposed a system that extracts bug reproduction steps directly from Android bug reports, facilitating automated debugging. Similarly, Li et al. \cite{10.1145/3377813.3381355} developed a bug reproduction system that utilizes user reviews as a primary source of information. However, system-level concurrency bugs differ significantly from those found in Android applications. Hence, these techniques are not applicable for this kind of bugs. Furthermore, Li et al. \cite{10.1145/3377813.3381355} relied on user feedback rather than structured bug reports, making their approach less suited for concurrency-related issues in system-level applications.

\noindent
\textbf{Structured IR technique to find similarity between a bug report and source code.} Previous works\cite{structuredIR, 6100061} have explored the use of Structured Information Retrieval (IR) techniques to determine the similarity between bug reports and source code. For instance, Saha et al. introduced BLUiR\cite{structuredIR}, a system that employs Structured IR to identify the most relevant files in the source code associated with a bug report. Their study demonstrated that their approach outperforms basic IR techniques. The primary objective of their work is to localize the buggy file within the source code. In contrast, our work focuses on a different goal. We aim to localize the system calls that are in buggy interleaving.

\noindent
\textbf{IR-Based Techniques to Localize Bugs.} Several information retrieval (IR)-based techniques have been proposed for bug localization. BugLocator \cite{6227210} leverages IR techniques to identify relevant source code files based on bug reports. Youm et al. introduced NAMe \cite{YOUM2017177} and BLIA \cite{7467300}, which enhance bug localization by incorporating change history alongside bug reports. BRTracer \cite{6976084} employs segmentation and stack-trace analysis to improve localization accuracy. AmaLgam \cite{10.1145/2597008.2597148} integrates version history, similar reports, and bug report structures to refine bug localization results. Locus \cite{7582764} further improves localization by analyzing software changes at a finer granularity than files, providing important contextual information for bug-fixing. Additionally, BLIZZARD \cite{10.1145/3236024.3236065} utilizes context-aware query reformulation to enhance IR-based bug localization. While these approaches effectively localize bugs, none are designed for automatic bug reproduction. Furthermore, they do not specifically target system-level concurrency bugs. Unlike \Name{}, the existing techniques do not account for system call names or failure-inducing inputs, which are crucial for reproducing System-level concurrency failures.

\noindent
\textbf{Extracting test cases from bug reports.} As there is a lexical mismatch between the bug report and technical language, many researchers use NLP techniques \cite{10.1145/3238147.3238204, recdroid}, and LLM \cite{hasan2025llput, 10.1145/3650212.3680383} to extract information from bug reports. For instance, PerfLearner\cite{10.1145/3238147.3238204} focuses on extracting information from bug reports to generate test frames, which then require manual conversion into concrete test cases. Another technique, Yakusu\cite{10.1145/3213846.3213869}, combines program analysis and NLP methods to create executable test cases from bug reports, primarily designed for Android mobile apps. However, it's worth noting that none of these approaches specifically target system-level concurrency bug reports. To the best of our knowledge, there has been no prior research conducted on the direct reproduction of bugs from system-level concurrency bug reports.

%% file: conclusion.tex
\section{Conclusion}
We have presented \Name{}, the first automated tool to reproduce system-level concurrency bugs from bug reports. \Name{} first automatically extracts relevant system call names from bug reports by combining natural language processing, data mining, and structured information retrieval techniques. Then \Name{} locates those system calls in the source code by using static analysis of the code. \Name{} generates input from the bug report by using information retrieval, regular expression matching, and the category-partition method. Finally, \Name{} uses the extracted input and interleaving information to reproduce the bug by performing dynamic binary instrumentation of the source code. The experimental results on 19 real-world subjects show that \Name{} is effective and efficient in reproducing system-level concurrency bugs from bug reports.